\documentclass[aps,prd,reprint,superscriptaddress,nofootinbib]{revtex4-1}
\usepackage{amssymb,amsfonts,amsmath,mathtools,hyperref,dsfont}
\usepackage[utf8]{inputenc}
\usepackage{enumitem}
\usepackage{bm}
\allowdisplaybreaks

\makeatletter
\def\namedlabel#1#2{\begingroup
    \normalfont #2\!%
    \def\@currentlabel{#2}%
    \phantomsection\label{#1}\endgroup
}
\makeatother

\def\eqn#1{eq.~\eqref{#1}}
\def\Eqn#1{Eq.~\eqref{#1}}
\def\eqns#1#2{eqs.~\eqref{#1} and~\eqref{#2}}

\def\sec#1{section~{\ref{#1}}}

\def\tab#1{table~{\ref{#1}}}

\def\sec#1{section~{\ref{#1}}}

\def\app#1{appendix~{\ref{#1}}}

\def\foot#1{footnote~{\ref{#1}}}

\def\bra#1{\langle #1|}
\def\ket#1{|#1 \rangle}

\def\braket#1{\langle #1 \rangle}
\def\la{\lambda}

\def\sgn{\operatorname*{sgn}}

\def\tr{\operatorname*{tr}}
\def\dim{\operatorname*{dim}}

\def\ie{{\it i.e.} }
\def\eg{{\it e.g.} }

\def\be{\begin{equation}}
\def\ee{\end{equation}}
\def\bea{\begin{eqnarray}}
\def\eea{\end{eqnarray}}  
\def\beal{\begin{equation}\begin{aligned}}
\def\eeal{\end{aligned}\end{equation}}
\def\nn{\nonumber}
\def\braket#1{\langle #1 \rangle}
\newcommand{\dd}{\mathrm{d}}

\begin{document} 
\title{Quantum anomalies from three-point on-shell bootstrap}

\author{Hiren Kakkad}
\email{hkakkad@shanghaitech.edu.cn}
\affiliation{Center for Fundamental Physics, School of Physical Science and Technology, ShanghaiTech University, 393 Middle Huaxia Road, Shanghai 201210, China}

\author{R\'emy Larue}
\email{ryrlarue@shanghaitech.edu.cn}
\affiliation{Center for Fundamental Physics, School of Physical Science and Technology, ShanghaiTech University, 393 Middle Huaxia Road, Shanghai 201210, China}

\author{Alexander Ochirov}
\email{ochirov@shanghaitech.edu.cn}
\affiliation{Center for Fundamental Physics, School of Physical Science and Technology, ShanghaiTech University, 393 Middle Huaxia Road, Shanghai 201210, China}

\begin{abstract}
We bootstrap quantum anomalies using on-shell techniques in the simplest setting: at three points. The necessary field-theoretic input includes the local or global classical symmetry, which the anomaly may break, and the symmetries of the effective action --- though not its explicit form. We use massless helicity spinors together with an on-shell representation of the discrete C, P, and T transformations. Our approach determines the Weyl, chiral (global and local), diffeomorphism, and Lorentz anomalies up to real constant prefactors and terms that vanish on shell. In particular, we recover gauge-anomaly cancellation conditions, as well as find that the Weyl anomaly should not involve Pontryagin densities.
\end{abstract}

\maketitle


\section{Introduction}
\label{sec:intro}

Quantum anomalies are among the most robust statements in field theory: a classical symmetry of the action can fail to survive quantization, and when it does,
there are significant formal and phenomenological consequences~\cite{Ball:1988xg,Bertlmann:1996xk,Fujikawa:2004cx}.
The most notorious example is the chiral anomaly, which governs the decay rate of the neutral pion through
the Adler--Bell--Jackiw anomaly~\cite{Adler:1969gk,Bell:1969ts}, is related to the topology of the gauge group via the index theorem~\cite{Atiyah:1968mp,Nielsen:1976hs,Nielsen:1977aw,Nielsen:1977qm,Christensen:1978md,Fujikawa:1979ay,Alvarez-Gaume:1983ict}, and puts strong constraints on the Standard Model (SM), beyond the SM, and string theory via the gauge-anomaly cancellation~\cite{Bouchiat:1972iq,Gross:1972pv,Green:1984sg} and 't Hooft anomaly matching~\cite{tHooft:1979rat}.
Another well-known example is the Weyl anomaly~\cite{Capper:1974ic}, which drives quantum vacuum effects, such as the Hawking radiation and the Casimir effect~\cite{Christensen:1977jc,Setare:2000ba,Setare:2000py}, constrains string compactifications~\cite{Polyakov:1981rd,Callan:1985ia,Fradkin:1985ys}, and plays a central role in holography~\cite{Henningson:1998gx}.

That anomalies are tightly constrained by symmetry is not new. 
The Wess--Zumino consistency conditions~\cite{Wess:1971yu} and their cohomological realization~\cite{Becchi:1975nq,Bonora:1983ff,Dubois-Violette:1985bnc} have long shown that the allowed anomalies are fixed by the gauge-group structure, largely independent of microscopic dynamics.
However, these approaches characterize the anomalies only off shell and cannot be applied purely on shell, in the sense of the modern scattering-amplitude bootstrap~\cite{Bern:1994zx,Bern:1994cg,Dixon:1996wi,Britto:2004nc,Britto:2004ap,Britto:2005fq,Britto:2005ha,Anastasiou:2006jv,Forde:2007mi,Benincasa:2007xk,Giele:2008ve,Badger:2013gxa,Badger:2015lda,Badger:2017jhb,Abreu:2018jgq,Badger:2018enw,Abreu:2018zmy}.
Recent works~\cite{Huang:2013vha,Chen:2014eva,AccettulliHuber:2021uoa,Alviani:2025msx} did probe gauge anomalies via a clash between unitarity and locality in 4- and higher-point scattering amplitudes, but without deciphering the anomaly structure itself.
Besides, global anomalies, which do not jeopardize unitarity nor locality, have remained elusive.
One may instead ask which 3-point on-shell structures are compatible with Lorentz invariance, discrete symmetries, and residual gauge freedom, and whether they can be determined without reference to an explicit action, a cohomological complex, or a regularization scheme.

In this paper, we show that anomalies in four-dimensional quantum field theories (QFT) do admit such an on-shell description.
Starting from anomalous Ward identities, we construct the most general 3-point form factors for the chiral, Weyl, diffeomorphism, and Lorentz anomalies consistent with (residual) gauge symmetries, discrete symmetries, and Bose symmetry.
More precisely, we define the anomalies field-theoretically, without specifying the form of the action, and then bootstrap them directly on shell.
Our basic assumptions are:
\begin{description}[font=\normalfont,leftmargin=0pt,itemsep=-1.5pt,topsep=1.5pt,after=\vspace{1.5pt},labelwidth=\widthof{(SYM):\!}]
\item[\namedlabel{ClassicalSymmetry}{(SYM)\;\!}] we only bootstrap anomalies for continuous transformations that are already classical symmetries;
\item[\namedlabel{EffectiveAction}{(EA)\;\!}]\:\!\!effective action $\Gamma$ encodes full quantum dynamics;
\item[\namedlabel{CPT}{(CPT)\;\!}] ${\cal CPT}$ and global Lorentz invariance of $\Gamma$;
\item[\namedlabel{CParity}{(CP)\;\!}] ${\cal C}$ and ${\cal P}$ invariance unless broken classically;
\item[\namedlabel{OneLoop}{(1L)\;\!}] we only bootstrap anomalies at one loop and hence use one-loop additivity in the particle content;
\item[\namedlabel{Covariant}{(COV)\;\!}] anomalies may be expressed in a covariant form.
\end{description}
\vspace{-0.4cm}
As a corollary, since covariant anomalies must be built out of gauge-covariant tensors, we also have
\begin{description}[font=\normalfont,leftmargin=0pt,itemsep=-1.5pt,topsep=1.5pt,after=\vspace{1.5pt},labelwidth=\widthof{(SYM):\!}]
\item[\namedlabel{Polynomial}{(POL)\;\!}] on shell, covariant anomalies must correspond to polynomials in helicity spinors at lowest order.
\end{description}
A detailed argument for \ref{Polynomial} will be given in \sec{sec:GaugeAnomaly}.

In this way, we will constrain the admissible spinor and tensor structures without any loop integration for each anomaly up to a real numerical coefficient, which depends on the microscopic details of the theory and is assumed to be nonzero.
Our bootstrap reproduces the known gauge and gravitational anomalies, including the mixed sectors. 
The anomaly-cancellation conditions emerge from the bootstrap as a symmetry-preserving constraint, rather than from a specific Lagrangian.
The resulting 3-point structures provide the seed from which the full anomaly can be reconstructed.
Our results prove that anomalies, like amplitudes, admit a description in terms of consistency rather than explicit loop calculations.

\section{Gauge anomaly}
\label{sec:GaugeAnomaly}

We start by bootstrapping anomalies in unitary gauge theories.
The gauge transformation $A_\mu \to U A_\mu U^{-1} + iU \partial_\mu U^{-1}/g$, with $U(x) = e^{ig\xi^a(x) t^a}$, has the linear part
\begin{equation}
\delta^\text{g}_\xi A^a_\mu = \partial_\mu \xi^a - g f^{abc} \xi^b A^c_\mu \eqqcolon D_\mu^\text{adj} \xi^a .
\label{eq:GaugeTransformation}
\end{equation}
As per assumption~\ref{EffectiveAction}, a certain effective-action functional $\Gamma[A,\varphi]$ fully describes the quantum dynamics of a gauge theory with matter transforming according to its, presently unspecified, gauge-group representation~$r$:\footnote{The gauge-group generators in representation~$r$ may be chosen hermitian and normalized to $\tr(t_r^a t_r^b) = C_r\delta^{ab}$, with $C_\text{F} = 1/2$ in the fundamental representation and $C_\text{A}=N$ in the adjoint of ${\rm SU}(N)$.
The anticommutation relations $\{t_r^a,t_r^b\}=d_r^{abc}t^c/C_r +\tilde{C}_r\delta^{ab}$ depend on~$r$, as opposed to $[t_r^a,t_r^b]=if^{abc}t_r^c$.
For ${\rm SU}(N)$ we have the standard relationship $\tilde{C}_r=2C_r/\dim(r)$, whereas for the abelian case of U(1), we choose $t^1=1$, $d^{111}=2$ and $\tilde{C}=0$ to avoid treating the abelian and nonabelian cases separately.
}
$\varphi \to U_r \varphi = \varphi + ig \xi^a t_r^a \varphi + {\cal O}(\xi^2)$.
Its gauge variation is
\begin{equation}
\delta^\text{g}_\xi \Gamma[A,\varphi] =\!\int\!\dd^4x \bigg[ \frac{\delta\Gamma}{\delta A^a_\mu\!} D_\mu^\text{adj} \xi^a + ig \xi^a \frac{\delta\Gamma}{\delta\varphi} t_r^a \varphi \bigg] .
\label{eq:GaugeVariation}
\end{equation}
The gauge-anomaly functional may be defined as
\begin{equation}\!\!
{\cal A}_\text{cons}^c(z)[A,\varphi] \coloneqq -\frac{\delta(\delta^\text{g}_\xi \Gamma)}{\delta\xi^c(z)} 
 = D_\mu^\text{adj} \frac{\delta\Gamma}{\delta A^c_\mu\!} - ig \frac{\delta\Gamma}{\delta\varphi} t_r^c \varphi .\!
\label{eq:GaugeAnomaly}
\end{equation}
The anomaly defined in this way from the effective action is the so-called consistent anomaly~\cite{Wess:1971yu}.
Consistent anomalies are generally not gauge-covariant and are thus harder to bootstrap.
There are, however, ambiguities in the definition of the composite operators for the current, which let us define the covariant anomaly~\cite{Bardeen:1984pm,Bertlmann:1996xk}
\begin{equation}
{\cal A}^c(z)[A,\varphi] \coloneqq {\cal A}_\text{cons}^c(z)[A,\varphi] + D_\mu^\text{adj} {\cal P}^{c\mu}(z)[A,\varphi] .
\label{eq:GaugeAnomalyCov}
\end{equation}
Here ${\cal P}^\mu$ is the local Bardeen--Zumino (BZ) polynomial in the fields.
We will not require its explicit form, but will only need to assume that it exists by~\ref{Covariant} and has the same mass dimension and symmetries as $\delta\Gamma/\delta A^c_\mu$.

In order to probe the covariant-anomaly functional~\eqref{eq:GaugeAnomalyCov} at the lowest nontrivial order, we further differentiate it twice and obtain the anomaly's 3-point function
\begin{equation} \begin{aligned}
{\cal A}_{\la\mu}^{abc}(x,y,z) \coloneqq\,& \frac{\partial~}{\partial z^\nu} \frac{\delta^2 \big[\delta{\Gamma/\delta A^{c\nu}(z)} + {\cal P}^{c\nu}(z)\big]}{\delta A^{a\la}(x) \delta A^{b\mu}(y)} \bigg|_{A=0} \\
 \eqqcolon &\,\frac{\partial~}{\partial z^\nu} \frac{\delta^3 \Gamma[A,0]}{\delta A^{a\la}(x) \delta A^{b\mu}(y) \delta A_\nu^c(z)} \bigg|_{A=0}^\text{cov} .
\label{eq:GaugeAnomaly3pt}
\end{aligned} \end{equation}
Here we have defined the shorthand ``cov'' for ``the covariant form of'', which lets us read off various properties of the anomaly that are unaltered by the addition of the BZ polynomial.
Due to our classical-symmetry assumption~\ref{ClassicalSymmetry}, the right-hand side comes exclusively from loop corrections.
In momentum space, it factorizes onto the momentum-conserving delta function.
We may then choose to put all three legs on shell, contract the vector indices with the appropriate polarizations and thus define the on-shell ``anomaly amplitude'' as\footnote{We discard the bubbles to obtain the second line in \eqn{eq:GaugeAnomaly3pt}, because they do not contribute to on-shell anomaly amplitudes.
}
\begin{align}
\label{eq:GaugeAnomaly3ptOnShell}
(2\pi)^4 & \delta^4(p_1+p_2+p_3)\,{\cal A}^{abc}(1^{h_1}\!,2^{h_2}\!,3) \\* \coloneqq &\,
{-}i\varepsilon_1^\la \varepsilon_2^\mu p_3^\nu \frac{\delta^3 \Gamma[A,0]}{\delta A^{a\la}(p_1) \delta A^{b\mu}(p_2) \delta A^{c\nu}(p_3)} \bigg|_{A=0,\,p_i^2=0}^\text{cov}\!. \nn
\end{align}

\Eqn{eq:GaugeAnomaly3ptOnShell} is what we aim to bootstrap.
It is well known~\cite{Witten:2003nn,Britto:2004nc,Britto:2005fq,Benincasa:2007xk} that the 3-point on-shell kinematics is overconstrained to have real-valued solutions, but has two distinct complex branches.
On the so-called MHV branch, the right-handed helicity spinors\footnote{Recall that massless ${\rm SL}(2,\mathbb{C})$ helicity spinors are defined via factorization of the $2\times2$ momentum matrices~\cite{Berends:1981rb,DeCausmaecker:1981jtq,Gunion:1985vca,Kleiss:1985yh,Xu:1986xb,Gastmans:1990xh}
\begin{equation*} \begin{aligned}
p_{\alpha\dot\beta} \coloneqq p_\mu \sigma_{\alpha\dot\beta}^\mu \eqqcolon \ket{p}_\alpha [p|_{\dot\beta} , \\
p^{\dot\alpha\beta} \coloneqq p^\mu \bar\sigma_\mu^{\dot\alpha\beta} \overset{\substack{\vspace{-8pt}\\ \Updownarrow}}{\eqqcolon} 
|p]^{\dot\alpha} \bra{p}^\beta .
\end{aligned} \end{equation*}
Here $\sigma^\mu = (1,\sigma^1,\sigma^2,\sigma^3) = \bar\sigma_\mu$ are the Pauli matrices with different chiral-index structures.
These indices are raised and lowered by the Levi-Civita tensors obeying
$\epsilon^{\alpha\beta}=-\epsilon_{\alpha\beta}=\smash{\epsilon^{\dot\alpha\dot\beta}=-\epsilon_{\dot\alpha\dot\beta}}$.
\label{foot:SpinorHelicity}
}
are all proportional to each other, $\smash{|1]^{\dot\alpha}\!\propto |2]^{\dot\alpha}\!\propto |3]^{\dot\alpha}}$.
$\smash{\overline{\text{MHV}}}$ means $\smash{\ket{1}_\alpha\!\propto \ket{2}_\alpha\!\propto \ket{3}_\alpha}$ for the left-handed spinors.
As all of the momentum invariants $p_i \cdot p_j = \braket{ij}[ji]/2$ vanish, nontrivial gauge-invariant expressions may only be constructed out of left-handed spinor products $\braket{ij}\neq0$ on the MHV branch and the right-handed $[ij]\neq0$ on the $\smash{\overline{\text{MHV}}}$ branch.
A straightforward analysis~\cite{Benincasa:2007xk,McGady:2013sga} of available options nails down the only two possible expressions according to the helicities~$h_i$ of each external leg:%
\begin{subequations} \begin{align}
\label{eq:General3ptMHV}
\text{MHV}\!: & \quad
\braket{12}^{h_3-h_1-h_2} \braket{23}^{h_1-h_2-h_3} \braket{31}^{h_2-h_3-h_1} , \\*
\label{eq:General3ptMHVbar}
\overline{\text{MHV}}\!: & \quad
[12]^{h_1+h_2-h_3} [23]^{h_2+h_3-h_1} [31]^{h_3+h_1-h_2} ,
\end{align} \label{eq:General3pt}%
\end{subequations}
which have mass dimensions~$\mp(h_1+h_2+h_3)$.
Consequently, demanding finiteness on the real MHV$\,\cap\,\overline{\text{MHV}}$ kinematics
forces us to pick the MHV solution whenever $h_1+h_2+h_3 \leq 0$ and the $\smash{\overline{\text{MHV}}}$ one for $h_1+h_2+h_3 \geq 0$.
(All helicities are understood for incoming momenta.)

For a concrete example, consider the $(-,-,+)$ 3-gluon amplitude $\propto \braket{12}^3 \braket{23}^{-1} \braket{31}^{-1}$.
It is clearly nonpolynomial.
Such spinor-helicity nonpolynomiality may originate from local field-theoretic (EFT) operators, whenever the gauge-dependent spinorial denominators inside the standard polarization vectors (see \eg \cite{Dixon:1996wi})
\begin{equation}
\varepsilon_{p,+}^\mu(q) =  \frac{1}{\sqrt{2}} \frac{\bra{q}\sigma^\mu|p]}{\braket{qp}} , \qquad
\varepsilon_{p,-}^\mu(q) = -\frac{1}{\sqrt{2}} \frac{[q|\bar\sigma^\mu\ket{p}}{[qp]}
\label{eq:PolVectors}
\end{equation}
do not cancel against the rest of the expressions.
In 3-gluon (and 3-graviton) amplitudes, this noncancellation is due to the cubic vertices not being covariant on their own, even at the linearized level: instead they are tied by covariance to lower points (\ie the free action)~\cite{Arkani-Hamed:2008bsc}.
On the other hand, this will never happen to any gauge-invariant effective-field-theoretic (EFT) operator that starts contributing at 3 points.
Indeed, its lowest-order truncation has to be invariant under the linearized gauge transformations, such as $A_\mu \to A_\mu+\partial_\mu\xi$.
For the polarization vectors~\eqref{eq:PolVectors}, the linearized gauge freedom lies in the reference-spinor dependence:
\begin{equation}
\varepsilon_{p,+}^\mu(q') = \varepsilon_{p,+}^\mu(q) + \tfrac{\sqrt{2}\braket{q'\!q}}{\braket{pq'}\braket{qp}} p^\mu .
\end{equation}
So the gauge-dependent denominators $\braket{qp}$ and $[qp]$ must cancel against the $q$-dependence in the numerator, as it happens \eg in \eqn{eq:F2onshell} below.
Therefore, we may view the polynomiality~\ref{Polynomial} as a corollary of \ref{Covariant}.

Let us now constrain the on-shell gauge anomaly~\eqref{eq:GaugeAnomaly3ptOnShell}, which has massless spins~$(1,1,0)$ and (mass) dimension~2, as explained \app{app:Effective}.
First, notice that the borderline cases $h_1+h_2+h_3=0$ are automatically ruled out by the polynomiality assumption~\ref{Polynomial}.\footnote{For instance, the helicities $h_1=-h_2=+1$ imply $\braket{23}^2/\braket{13}^2$ on the MHV branch and $[13]^2/[23]^2$ on the $\smash{\overline{\text{MHV}}}$ one, which are dimensionless instances of \eqn{eq:General3pt}.
They can thus be set to zero by dimension counting, assuming the absence of positively dimensionful coupling constants --- which notably leaves room for higher-dimensional EFT couplings.
Such 3-point objects are also known to lead to higher-point inconsistencies~\cite{McGady:2013sga}, which would contradict our covariance assumption~\ref{Covariant}.
These further observations support our polynomial assumption~\ref{Polynomial}.
\label{foot:DimensionCounting}
}
The remaining $(\mp1,\mp1,0)$ helicity configurations correspond to the unique dimension-2 MHV and $\smash{\overline{\text{MHV}}}$ expressions~\eqref{eq:General3pt}, namely $\braket{12}^2$ and $[12]^2$, respectively.
At this point, these on-shell answers still have undetermined dimensionless prefactors, which may be two different functions of the gauge coupling~$g$ times the color factor.
At 3 points, this color factor can only be a combination of the fully antisymmetric structure constants $f^{abc}=-i\tr[t_r^a,t_r^b]t_r^c/C_r\in\mathbb{R}$ and the fully symmetric $d_r^{abc}\coloneqq\tr\{t_r^a,t_r^b\}t_r^c\in\mathbb{R}$, where the generators~$t_r^a$ are understood to be in a generic representation~$r$ of the gauge group.
However, the Bose symmetry between legs~1 and~2 in combination with the kinematic symmetry $1\leftrightarrow2$ of the squared spinor brackets implies that the color factor must be a linear combination of $d^{abc}_r$ only.
For $d_r^{abc}$ to be nonzero, however, the representation~$r$ cannot be real (\ie with all antisymmetric generators).
In particular, it cannot be adjoint, so we are naturally led to consider matter particles (\ie not gauge bosons) in a complex representation~$r$ propagating in a loop.
Therefore, only such matter particles contribute to the on-shell anomaly and have the form\footnote{
The gauge coupling~$g$ in \eqn{eq:GaugeAnomalyBrakets} could be rescaled into a coupling specific to the particle in the loop, see the final answer~\eqref{eq:GaugeAnomalyFinal}.
}
\begin{equation}
{\cal A}^{abc}(1^{h_1}\!,2^{h_2}\!,3)=g^3 d_r^{abc}\times\left\{\!
\begin{smallmatrix}
   C^- \braket{12}^2 , & h_1=h_2=-1 , \\
   C^+ [12]^2 , & h_1=h_2=+1 , \\
   0 , & \text{otherwise} ,
\end{smallmatrix}\right.
\label{eq:GaugeAnomalyBrakets}
\end{equation}
where $C^\pm$ are dimensionless constant parameters.\footnote{Note that in principle one could include EFT corrections involving an intrinsic scale $\Lambda$.
However, the only dimensionless polynomial invariants that one may write without spoiling the helicity weights are $\braket{ij}[ji]/\Lambda^2$, which vanish on shell.
\label{foot:NonMinCoup}
}

To proceed to various concrete versions of the gauge anomaly, we need to consider different ${\cal C}$, ${\cal P}$, and ${\cal T}$ properties of the (so far generic) gauge field~$A_\mu$.
First, we assume it to be a proper vector field, which is odd under charge conjugation~${\cal C}$, see the first column of \tab{tab:VALR}, as is also its gauge parameter~$\xi$.
We use a representation of ${\cal C}$ at the on-shell level, which follows from its unitary action on the Fock space at the quantum-field level \cite{Kakkad:2026long}:
\be\!\!
U_{\cal C} A^\mu_{\text{V}i\bar\jmath} U_{\cal C}^{-1}\!= -A^\mu_{\text{V}j\bar\imath} ~\Rightarrow~
{\cal C} A_\text{V}^{a\mu} = -A_\text{V}^{a\mu} ,~ {\cal C} t^a_{i\bar\jmath} = t^a_{j\bar\imath} .
\label{eq:ChargeConjugationVector}
\ee
This action of ${\cal C}$ implies 
${\cal C}\:\!d_r^{abc} = d_r^{abc}$. Since ${\cal C}$ does not affect the spinors in any way, the anomaly amplitude~\eqref{eq:GaugeAnomalyBrakets} is invariant under $\cal C$. On the other hand, the ${\cal C}$-invariance assumption ${\cal C}\Gamma=\Gamma$ along with the vector-field transformation given in \tab{tab:VALR} implies ${\cal C A}^{abc} = -{\cal A}^{abc}$, as follows from the definition~\eqref{eq:GaugeAnomaly3ptOnShell}.
This contradiction means that the anomaly is forbidden in the purely vectorial case.\footnote{This argument based on ${\cal C}$ invariance may be regarded as a generalized version of Furry's theorem.
}

\begin{table}[t]
\centering
\begin{tabular}{c||c|c|c|c|c|c}
             &  $V^\mu_{i\bar\jmath}$ &  $A^\mu_{i\bar\jmath}$ &  $L^\mu_{i\bar\jmath}$ &  $R^\mu_{i\bar\jmath}$ & $\partial_\mu$ & $\gamma_5$ \\ \hline \hline
${\cal C}$   & $-V^\mu_{j\bar\imath}$ & $+A^\mu_{j\bar\imath}$ & $-\frac{g_\text{R}}{g_\text{L}} R^\mu_{j\bar\imath}$ & $-\frac{g_\text{L}}{g_\text{R}} L^\mu_{j\bar\imath}$ & $+\partial_\mu$ & $-\gamma_5^\top\phantom{\big|}\!$ \\
${\cal P}$   & $+V_{\mu i\bar\jmath}$ & $-A_{\mu i\bar\jmath}$ & $+\frac{g_\text{R}}{g_\text{L}} R_{\mu i\bar\jmath}$ & $+\frac{g_\text{L}}{g_\text{R}} L_{\mu i\bar\jmath}$ & $+\partial^\mu$ & $-\gamma_5$ \\
${\cal T}$   & $+V_{\mu i\bar\jmath}$ & $+A_{\mu i\bar\jmath}$ & $+L_{\mu i\bar\jmath}$ & $+R_{\mu i\bar\jmath}$ & $-\partial^\mu$ & $+\gamma_5$ \\
${\cal CPT}$ & $-V^\mu_{j\bar\imath}$ & $-A^\mu_{j\bar\imath}$ & $-L^\mu_{j\bar\imath}$ & $-R^\mu_{j\bar\imath}$ & $-\partial_\mu$ & $+\gamma_5^\top$ \\
\end{tabular}
\caption{Charge conjugation~${\cal C}$, parity~${\cal P}$, and time reversal~${\cal T}$ for different gauge fields.
For legibility, we use a condensed notation for $A_\text{V}^\mu=V^\mu$, $A_\text{A}^\mu=A^\mu$, $A_\text{L}^\mu=L^\mu$, $A_\text{R}^\mu=R^\mu$.
}
\label{tab:VALR}
\end{table}

To consider other gauge-field configurations, we dress the coefficients~$C^\pm_{\pi_1\pi_2\pi_3}$ in \eqn{eq:GaugeAnomalyBrakets} with additional labels that encode parity of each leg: $\pi_j \in \{\text{V},\text{A}\}$ for ``vector'' and ``axial vector'', respectively; \eg we just showed $C^\pm_\text{VVV}=0$.
No matter how exactly the matter fields~$\varphi$ couple to the gauge field~$A$, gauge invariance requires the minimal coupling terms to complement the kinetic term of~$\varphi$ via the covariant derivative defined as
\begin{equation}\!\!
D^\mu \varphi \coloneqq \partial^\mu \varphi - ig A^{a\mu} t_r^a \varphi , \quad
D^\mu \widetilde\varphi \coloneqq \partial^\mu \widetilde\varphi + ig \widetilde\varphi\,t_r^a A^{a\mu}\!.\!
\label{eq:CovariantDerivative}
\end{equation}
Here $\widetilde\varphi$ is the dual field with the opposite charge.
${\cal C}$ invariance makes these derivatives transform into each other in the same way as the corresponding fields, so $A^\mu$ must be a vector (and not axial vector).
This immediately prevents gauge anomalies to arise from spin-0 and spin-1 matter.

Let us now focus on the fermionic case, in which we can preserve $\cal C$ (and ${\cal P}$) by combining an axial gauge field with~$\gamma_5$.
The standard conjugation rule for a Dirac-spinor field reads:
\begin{equation}\!\!\!\!
\begin{aligned}
U_{\cal C} \Psi_j U_{\cal C}^{-1} & = \varkappa\,\overline\Psi_{\bar\jmath}\,C_\text{D} , 
\end{aligned} \quad
C_\text{D} \coloneqq -i\gamma^0 \gamma^2
 = \Big(\!\begin{smallmatrix}\!\epsilon^{\alpha\beta}\!& 0 \\
                               0 &\!\epsilon_{\dot{\alpha}\dot{\beta}}\!
           \end{smallmatrix}\!\Big)\!
\label{eq:ChargeConjugationDirac}
\end{equation}
(where $\varkappa$ is an arbitrary constant phase factor), which in terms of the Weyl-field constituents translates to
\be
\Psi_j = \Big(\!
\begin{smallmatrix} \chi_{\alpha j} \\
                    \tilde\eta^{\dot\alpha}_j \end{smallmatrix}\!\Big)
\qquad \Rightarrow \qquad \left\{
\begin{aligned}
U_{\cal C}\,\chi_{\alpha j}\,U_{\cal C}^{-1} & = \varkappa\,\eta_{\alpha\bar\jmath} , \\
U_{\cal C}\,\tilde\eta^{\dot\alpha}_j\,U_{\cal C}^{-1} & = \varkappa\,\tilde\chi^{\dot\alpha}_{\bar\jmath} .
\end{aligned}\right.
\label{eq:ChargeConjugationWeyl}
\ee
It is then easy to derive from \tab{tab:VALR} that the conjugation rule $U_{\cal C}\,D_\mu \Psi\,U_{\cal C}^{-1} = \varkappa\,D_\mu\overline\Psi\,C_\text{D}$ holds for a covariant derivative that may include any linear combination of vector and axial gauge fields:
\be
D^\mu \Psi = [\partial^\mu - igA_\text{V}^\mu - ig'\!A_\text{A}^\mu\gamma_5] \Psi .
\label{eq:CovariantDerivativeVA}
\ee
It may be conveniently rewritten for Weyl fermions as
\begin{subequations} \begin{align}\!\!\!
D^\mu \chi = [\partial^\mu\!- ig A_\text{V}^\mu + ig'\!A_\text{A}^\mu] \chi
& \eqqcolon [\partial^\mu\!- ig_\text{L} A_\text{L}^{a\mu} t_r^a] \chi , \\\!\!\!
D^\mu \tilde\eta = [\partial^\mu\!- ig A_\text{V}^\mu - ig'\!A_\text{A}^\mu] \tilde\eta
& \eqqcolon [\partial^\mu\!- ig_\text{R} A_\text{R}^{a\mu}t_r^a] \tilde\eta ,
\end{align} \label{eq:CovariantDerivative2}%
\end{subequations}
in terms of the ``left'' and ``right'' gauge fields.
They gauge two independent transformations $U_\text{L}(x) = e^{ig_\text{L}\xi_\text{L}^a(x) t_r^a}$ and $U_\text{R}(x) = e^{ig_\text{R}\xi_\text{R}^a(x) t_r^a}$.
The mass term $\bar{\Psi}\Psi = \eta^\alpha \chi_\alpha + \tilde\chi_{\dot\alpha} \tilde\eta^{\dot\alpha}$, however, is not invariant under such transformations, so massive Dirac fermions do not produce gauge anomalies.\footnote{We only consider a bare mass in our discussion.
Massive chiral fermions can be achieved using spontaneous symmetry breaking~\cite{Higgs:1964pj} and hence produce an anomaly (see \eg~\cite{,Quevillon:2021sfz,Filoche:2022dxl}).
}
Moreover, since the constant prefactors in \eqn{eq:GaugeAnomalyBrakets} are dimensionless, massless (vectorlike) Dirac fermions do not either.
So we have determined that such anomalies require chiral massless fermions that are independently gauged at the classical level.

We may now regard the left and right gauged matter fields as independent, so their representations in \eqn{eq:CovariantDerivative2} and $U_\text{L/R}(x)$ do not have to match up and should be distinguished as~$r_\text{L/R}$.
Owing to one-loop additivity~\ref{OneLoop}, we can focus on the case of one left- and one right-handed fermion.
Since the left and right currents going through the loop are completely independent,\footnote{At the level of minimal coupling, the mixed L/R interactions $[\tilde\eta|A_\text{L/R}\ket{\chi}$ and $\bra{\chi}A_\text{L/R}|\tilde\eta]$ would imply the chiral-field mixing already inside the free action and should thus be forbidden for a diagonal choice of free fields.
Mixed L/R anomaly contributions could, however, be generated by nonminimal couplings, but as emphasized in \foot{foot:NonMinCoup}, the on-shell 3-point bootstrap is oblivious to such corrections.
}
all mixed anomaly contributions like ${\cal A}_\text{LLR}^{abc}$ vanish.
This leaves only two contributions ${\cal A}_\text{LLL}^{abc}$ and ${\cal A}_\text{RRR}^{abc}$ of the form~\eqref{eq:GaugeAnomalyBrakets} but involving $g_\text{L}^3 d_{r_\text{L}}^{abc} C_\text{LLL}^\pm$ and $g_\text{R}^3 d_{r_\text{R}}^{abc} C_\text{RRR}^\pm$, respectively.
The freedom to choose representations, however, still lets us pick the same representation $r_\text{L}=r_\text{R}=r$.
Then using~\ref{OneLoop}, we can project back to the vanishing purely vectorial Dirac-fermion anomaly ${\cal A}_\text{VVV}^{abc}$ via the definitions~\eqref{eq:GaugeAnomaly3ptOnShell} and~\eqref{eq:CovariantDerivative2}:
\beal
\label{eq:Weyl2Dirac}
\frac{g^3}{g_\text{L}^3\!} {\cal A}_\text{LLL}^{abc} + \frac{g^3}{g_\text{R}^3\!} {\cal A}_\text{RRR}^{abc} \bigg|_{r_\text{L}=r_\text{R}}\!\!\!\!\!\!\!\!\!\!\!= 0 ~\,\Rightarrow\,~ -C_\text{LLL}^\pm = C_\text{RRR}^\pm \eqqcolon C^\pm .
\eeal

\begin{table}[t]
\centering
\begin{tabular}{c||c|c|c|c}
    &~${\cal C}$ {\scriptsize (unitary)}~&~${\cal P}$ {\scriptsize (unitary)}~&~${\cal T}$ {\scriptsize (antiunitary)}~&~${\cal CPT}$ \\ \hline \hline
$t^a_{i\bar\jmath}$ & $t^a_{j\bar\imath}$~~& $t^a_{i\bar\jmath}$ {\scriptsize (inv.)} & $t^a_{i\bar\jmath}$ {\scriptsize (inv.)} & $t^a_{j\bar\imath}$ \\
$\ket{p}_\alpha$ & $\ket{p}_\alpha$~\:\!\!{\scriptsize(inv.)} & $|p]^{\dot\alpha}$ & $\bra{p}^\alpha$ & $-[p|_{\dot\alpha}$ \\
$|p]^{\dot\alpha}$ & $|p]^{\dot\alpha}$~{\scriptsize (inv.)} & $-\ket{p}_\alpha$ & $-[p|_{\dot\alpha}$ & $-\bra{p}^\alpha$ \\
$p^\mu$ & $p^\mu$~{\scriptsize (inv.)} & $p_\mu$ & $p_\mu$ & $~p^\mu$~{\scriptsize (inv.)}\!\\
$\varepsilon_{p,\pm}^\mu$ & $\varepsilon_{p,\pm}^\mu$~{\scriptsize (inv.)} & $\varepsilon^\mp_{p\mu}$ & $\varepsilon^\pm_{p\mu}$ & $\varepsilon_{p,\mp}^\mu$ \\
$i$ &~$i$~~~\;{\scriptsize (inv.)} &~$i$~~{\scriptsize (inv.)} & $-i$ & $-i$ \\
\end{tabular}
\caption{Action of ${\cal C}$, ${\cal P}$, and ${\cal T}$ on on-shell building blocks.
The conjugation rules shown for generators~$t^a_{r_\text{V}}$ are consistent with vector, axial, left, and right gauge fields.
A more detailed exposition of these conventions is provided in \app{app:PT}.
}
\label{tab:CPT}
\end{table}%

$C^\pm$ can be further related via parity in this ${\cal P}$-invariant setup.
Indeed, the assumption~\ref{CParity} together with the representation choice $r_{\text{L}}=r_{\text{R}}$ implies ${\cal P}\Gamma=\Gamma$.
Using this inside the definition~\eqref{eq:GaugeAnomaly3ptOnShell} for ${\cal A}_\text{LLL}^{abc}$ and ${\cal A}_\text{RRR}^{abc}$ (spelled out explicitly in \eqn{eq:GaugeAnomaly3ptChiral}), along with the action of $\cal P$ on the gauge fields from \tab{tab:VALR} and on the polarizations,
${\cal P}\varepsilon_{p,\pm}^\mu=\varepsilon^\mp_{p\mu}$, we obtain
\be\!
{\cal P A}_\text{LLL}^{abc}(1^-\!,2^-\!,3) = \tfrac{g_\text{L}^3}{g_\text{R}^3} {\cal A}_\text{RRR}^{abc}(1^+\!,2^+\!,3) = g_\text{L}^3 d_r^{abc} C^+ [12]^2\!.
\label{eq:GaugeAnomalyParity1}
\ee
On the other hand, we can use a purely on-shell action of parity on the kinematic building blocks, which is shown in \tab{tab:CPT} along with ${\cal C}$ and ${\cal T}$.
(More details about the spinor-helicity representation of ${\cal P}$ and ${\cal T}$ may be found in \app{app:PT}.)
Applying this transformation to the on-shell ansatz of the same anomaly amplitude leads to
\begin{equation}
{\cal P} \big\{{-}g_\text{L}^3 d_r^{abc} C^- \braket{12}^2 \big\}
 = -g_\text{L}^3 d_r^{abc} C^- [12]^2 .
\label{eq:GaugeAnomalyParity2}
\end{equation}
Comparing \eqns{eq:GaugeAnomalyParity1}{eq:GaugeAnomalyParity2}, we conclude that $-C^-=C^+\eqqcolon iC$.
The reality of $C$ may be seen in a similar fashion from ${\cal CPT}$ symmetry, which also maps between the MHV and $\smash{\overline{\text{MHV}}}$ sectors but without changing handedness.
In this way, we arrive at the following nonzero on-shell anomalies:
\beal\!\!\!\!\!\!
{\cal A}_\text{LLL}^{abc}(1^{h_1}\!\!,2^{h_2}\!\!,3) &
 = iC\!\!\!\!\!\!\sum_{\substack{\text{\scriptsize left Weyl}\\\text{\scriptsize fermions }\chi}}\!\!\!\!\!\!g_\chi^3 d_{r_\chi}^{abc}\!\times\!\!
\left\{\!
\begin{smallmatrix}
   \braket{12}^2\!, & h_1=h_2=-1 , \\
   -[12]^2\!, & h_1=h_2=+1 , \\
    0 , & \text{otherwise,} \\
\end{smallmatrix} \right.\!\!\!\!\!\!\!\\\!\!\!\!\!\!
{\cal A}_\text{RRR}^{abc}(1^{h_1}\!\!,2^{h_2}\!\!,3) &
 = iC\!\!\!\!\!\!\sum_{\substack{\text{\scriptsize right Weyl}\\\text{\scriptsize fermions }\tilde\eta}}\!\!\!\!\!\!g_{\tilde\eta}^3 d_{r_{\tilde\eta}}^{abc}\!\times\!\!
\left\{\!
\begin{smallmatrix}
   -\braket{12}^2\!, & h_1=h_2=-1 , \\
   [12]^2\!, & h_1=h_2=+1 , \\
    0 , & \text{otherwise.} \\
\end{smallmatrix} \right.\!\!\!\!\!\!\!
\label{eq:GaugeAnomalyFinal}%
\eeal
Here we have again allowed the matter particles to have their own representations and rescaled versions of the gauge coupling~$g$, such as different charges~$Q_j$ in the abelian case (where $t^1=1$, $d^{111} = 2$).
For a single gauge field $A_\text{L}=A_\text{R}$ coupling to Weyl fermions, the gauge-anomaly cancellation condition can be read off \eqn{eq:GaugeAnomalyFinal}:
\begin{equation}
\sum_{\chi\in\text{left}}\!g_\chi^3 d^{abc}_{r_\chi} -\!\!\sum_{\tilde\eta\in\text{right}}\!\!g_{\bar\eta}^3 d^{abc}_{r_{\tilde\eta}}=0.
\end{equation}

The on-shell answer~\eqref{eq:GaugeAnomalyFinal} may also be uplifted to the off-shell covariant gauge anomaly~\eqref{eq:GaugeVariation} of the form
\beal\!\!\!
\delta^\text{g}_\xi \Gamma\big|^{\text{cov}}\!= -\frac{C}{2}\!\int\!\dd^4x \sqrt{|g|} \Big[\!\sum_{\chi\,\in\,\text{\scriptsize left}}\!\!g_\chi^3 & \tr\!{}_{r_\chi}(\xi_\text{L} F^\text{L}_{\mu\nu} \widetilde{F}_\text{L}^{\mu\nu})\!\\
 -\!\!\!\sum_{\tilde\eta\,\in\,\text{\scriptsize right}}\!\!\!g_{\tilde\eta}^3 & \tr\!{}_{r_{\tilde\eta}}(\xi_\text{R} F^\text{R}_{\mu\nu} \widetilde{F}_\text{R}^{\mu\nu})\:\!\Big] ,\!
\label{eq:GaugeVariationAnswer}
\eeal
where the dual field strengths are $\widetilde{F}_{\mu\nu}\coloneqq\epsilon_{\mu\nu\sigma\tau} F^{\sigma\tau}/2$.
Indeed, it is easy to check that the corresponding gauge-invariant building blocks
$f_{p,\pm}^{\mu\nu} \coloneqq -2ip^{[\mu} \varepsilon_{p,\pm}^{\nu]}$ give\footnote{Round (square) brackets denote index (anti)symmetrization and include normalization by factorial denominators, such as $1/2!$.
}
\begin{subequations} \begin{align}
\label{eq:FFonshell}
\eta_{\la\nu} \eta_{\mu\rho} f_{p_1,\mp}^{\la\mu} f_{p_2,\mp}^{\nu\rho} = \Big\{\!
   \begin{smallmatrix}
   -\braket{12}^2 & \text{ for } (-,-) , \\
   -[12]^2 & \text{ for } (+,+) ,
   \end{smallmatrix} \\
\label{eq:*FFonshell}
\frac{1}{2} \epsilon_{\la\mu\nu\rho} f_{p_1,\mp}^{\la\mu} f_{p_2,\mp}^{\nu\rho} = \Big\{\!
   \begin{smallmatrix}
   i\braket{12}^2 & \text{ for } (-,-) , \\
   -i[12]^2 & \text{ for } (+,+) ,
   \end{smallmatrix}
\end{align} \label{eq:F2onshell}%
\end{subequations}
as well as zeros for the mixed-helicity configurations.
So the relative sign between the helicity configurations discerns between $\xi F\!\cdot\!F$ and $\xi F\!\cdot\!\widetilde{F}$ and determines the chiral character of the anomaly~\eqref{eq:GaugeVariationAnswer}.
The overall minus sign in \eqn{eq:GaugeVariationAnswer} is due to our original sign convention in \eqn{eq:GaugeAnomaly}.
Comparing with the literature~\cite{Fujikawa:1983bg,Bertlmann:1996xk,Fujikawa:2004cx}, we can of course set $C=(8\pi^2)^{-1}$,
which is the only numerical factor that seems to elude the on-shell constraints.

\section{Gauge anomaly: other sectors}
\label{sec:ChiralAnomalyGrav}

Let us extend the chiral-anomaly bootstrap to include gravity perturbations around the Minkowski metric, $g_{\mu\nu} = \eta_{\mu\nu} + \kappa h_{\mu\nu}$.
Starting from the effective action $\Gamma[g_{\mu\nu},A^a_\mu,\varphi]$,
we have new 3-point on-shell anomalies (of mass dimension 2) defined in full analogy with \eqn{eq:GaugeAnomaly3ptOnShell}:%
\begin{subequations} \begin{align}
\label{eq:GaugeGravAnomalyMixed3pt}
(2\pi)^4 \delta^4(p_1+p_2+p_3)\,{\cal A}_{gA}^{bc}(1^{h_1}\!,2^{h_2}\!,3) \coloneqq & \\*
   {-}i\kappa\,\varepsilon_1^{\mu\nu} \varepsilon_2^\rho\,p_3^\la \frac{\delta^3 \Gamma[g,A,0]}{\delta g^{\mu\nu}(p_1)\delta A^{b\rho}(p_2)\delta A^{c\la}(p_3)} & \bigg|_{g=\eta,\,A=0,\,p_i^2=0}^\text{cov}\!, \nn \\
\label{eq:GaugeGravAnomalyPure3pt}
(2\pi)^4 \delta^4(p_1+p_2+p_3)\,{\cal A}_{gg}^c(1^{h_1}\!,2^{h_2}\!,3) \coloneqq & \\*
   {-}i\kappa^2 \varepsilon_1^{\mu\nu} \varepsilon_2^{\rho\sigma} p_3^\la \frac{\delta^3 \Gamma[g,A,0]}{\delta g^{\mu\nu}(p_1)\delta g^{\rho\sigma}(p_2)\delta A^{c\la}(p_3)} & \bigg|_{g=\eta,\,A=0,\,p_i^2=0}^\text{cov}\!.\!\nn
\end{align} \label{eq:GaugeGravAnomaly3pt}%
\end{subequations}

In the mixed sector~\eqref{eq:GaugeGravAnomalyMixed3pt}, the on-shell bootstrap~\eqref{eq:General3pt} gives nonpolynomial amplitudes, both for the same- and mixed-helicity configurations, such as $\braket{12}^3\braket{31}/\braket{23}$ and $\braket{12}\braket{31}^3/\braket{23}^3$, respectively.
These are all ruled out by our assumption~\ref{Polynomial}.
Consequently, the mixed gauge-gravitational anomaly~\eqref{eq:GaugeGravAnomalyMixed3pt} vanishes.

In the pure-gravity sector~\eqref{eq:GaugeGravAnomalyPure3pt}, the bootstrap~\eqref{eq:General3pt} unsurprisingly produces a double copy (see \eg \cite{Bern:2022wqg}) of the pure gauge anomaly, namely ${\cal A}_{gg}$ is proportional to $\braket{12}^4$ for $h_1=h_2=-2$, to $[12]^4$ for $h_1=h_2=+2$, and the mixed-helicity cases are ruled out by assumption~\ref{Polynomial}.
Since there is a single colored leg, the color factor must be $\tr t^c_r$, which is only non-vanishing in the abelian case.
The ${\cal C}$-invariance assumption~\ref{CParity} again rules out the case of a single vector gauge field, which again forces us to consider chiral fermions in the loop that may be gauged separately by $A^\mu_\text{L}$ and $A^\mu_\text{R}$.
Projecting back to the vanishing ${\cal P}$-even case of a vectorially gauged Dirac field implies that the constant prefactors satisfy $C_\text{L}^\pm = -C_\text{R}^\pm$.
Parity, on the other hand, converts ${\cal P A}_{gg\text{L}}(1^-\!,2^-\!,3) = \smash{\tfrac{Q_\text{L}}{Q_\text{R}}} {\cal A}_{gg\text{R}}(1^+\!,2^+\!,3)$ and thus equates
\begin{equation}
Q_\text{L} C_\text{L}^- [12]^4
= {\cal P}\big\{ Q_\text{L} C_\text{L}^- \braket{12}^4 \big\}
= \tfrac{Q_\text{L}}{Q_\text{R}} Q_\text{R} C_\text{R}^+ [12]^4 .\!
\label{eq:GaugeGravAnomalyParity}
\end{equation}
Here and below, switching $g$ for $Q$ conveys the abelian-group restriction $t^1=1$.
So we have $C_\text{L}^\mp = C_\text{R}^\pm = \pm iC$, with $C$ real due to ${\cal CPT}$ invariance.
Therefore, the so-called mixed axial-gravitational anomalies~\cite{Delbourgo:1972xb,Eguchi:1976db,Alvarez-Gaume:1983ihn} are
\begin{subequations} \begin{align}\!\!\!
{\cal A}_{gg\text{L}}(1^{h_1}\!\!,2^{h_2}\!\!,3) &
 = iC\kappa^2\!\!\!\sum_{\chi\,\in\,\text{\scriptsize left}}\!\!\!Q_\chi\!\times\!
\left\{\!
\begin{smallmatrix}
   \braket{12}^4\!,\!& h_1=h_2=-2 , \\
   \!-[12]^4\!,\!& h_1=h_2=+2 , \\
    0 , & \text{otherwise,} \\
\end{smallmatrix} \right.\!\\\!\!\!
{\cal A}_{gg\text{R}}(1^{h_1}\!\!,2^{h_2}\!\!,3) &
 = iC\kappa^2\!\!\!\!\sum_{\tilde\eta\,\in\,\text{\scriptsize right}}\!\!\!\!Q_{\tilde\eta}\!\times\!
\left\{\!
\begin{smallmatrix}
   \!-\braket{12}^4\!,\!& h_1=h_2=-2 , \\
   [12]^4\!,\!& h_1=h_2=+2 , \\
    0 , & \text{otherwise.} \\
\end{smallmatrix} \right.\!\!
\end{align} \label{eq:GaugeGravAnomalyFinal}%
\end{subequations}
The corresponding axial-gravitational anomaly cancellation condition may be read off directly from above as
\begin{equation}
\sum_{\chi\in\text{left}}\!Q_\chi-\!\!\sum_{\tilde\eta\in\text{right}}\!\!Q_{\tilde\eta} = 0 .
\label{eq:axialgravCancellation}
\end{equation}
Since the higher-point anomaly amplitudes are tied to the 3-point one by diffeomorphism invariance, we find the covariant off-shell gauge variation (in addition to \eqn{eq:GaugeVariationAnswer})
\begin{equation}
\delta'^{\text{g}}_\xi \Gamma\big|^{\text{cov}}\!= C\!\!\int\!\!\dd^4x \sqrt{|g|} R_{\la\mu\nu\rho} \widetilde{R}^{\la\mu\nu\rho} \big[ \xi_\text{L}\!\!\!\!\sum_{\chi\,\in\,\text{\scriptsize left}}\!\!\!\!Q_\chi
- \xi_\text{R}\!\!\!\!\!\sum_{\tilde\eta\,\in\,\text{\scriptsize right}}\!\!\!\!\!Q_{\tilde\eta} \big] .
\label{eq:GaugeGravVariationAnswer}
\end{equation}
Here $\widetilde{R}_{\la\mu\nu\rho} \coloneqq \sqrt{|g|} \epsilon_{\la\mu\sigma\tau} R^{\sigma\tau}{}_{\nu\rho}/2$ is the dual Riemann tensor, which is required for the relative signs in \eqn{eq:GaugeGravAnomalyFinal}.

\section{Diffeomorphism anomaly}
\label{sec:DiffeoAnomaly}

We consider the same effective action in curved spacetime $\Gamma[g_{\mu\nu},A^a_\mu,\varphi]$.
Such an action should be invariant under general coordinate transformations, which can be represented by the diffeomorphism group.
Under an infinitesimal diffeomorphism transformation $x^\mu \to x^\mu-\xi^\mu$, the fields transform along their Lie derivatives:
\begin{equation}\!\!\!
\delta^\text{d}_\xi g_{\mu\nu} =\nabla_\mu\xi_\nu+\nabla_\nu\xi_\mu ,\!\quad
\delta^\text{d}_\xi A^a_\mu =\xi^\nu \nabla_\nu A^a_\mu+A^a_\nu\nabla_\mu\xi^\nu ,
\label{eq:DiffeoTransformation}
\end{equation}
where $\nabla$ is the spacetime-covariant derivative that ignores gauge indices.
The vacuum expectation value (VEV) of the energy-momentum tensor (EMT), defined by
$\sqrt{|g|} \braket{T_{\mu\nu}} \coloneqq 2\delta\Gamma/\delta g^{\mu\nu} = -2g_{\mu\rho} g_{\nu\sigma} \delta\Gamma/\delta g_{\rho\sigma}$,
allows us to express the effective-action variation as 
\begin{align}
\delta^\text{d}_\xi \Gamma\!=\!\!\int\!\!\dd^4x \bigg\{ & \xi^\mu\!\bigg[\:\!\!\sqrt{|g|}\nabla^\la \braket{T_{\la\mu}}
 -\!\sqrt{|g|} A^c_\mu {\cal A}_{\text{cons}}^c\!+ F^c_{\mu\nu} \frac{\delta\Gamma}{\delta A^c_\nu}\:\!\!\bigg] \nn \\* &
 + \big[ \delta^\text{d}_\xi \varphi - (\xi\!\cdot\!A^c) ig t^c_r \varphi \big] \frac{\delta\Gamma}{\delta\varphi} \bigg\} .
\label{eq:deltaDiffGamma}
\end{align}
Here ${\cal A}_{\text{cons}}^c$ is the local gauge-anomaly functional~\eqref{eq:GaugeAnomaly}.
Analogously, we define the diffeomorphism-anomaly functional
${\cal A}_{\text{cons}}^\mu(x)\coloneqq -\delta(\delta^\text{d}_\xi \Gamma)/\delta\xi_\mu(x) = 2\nabla_\la \delta\Gamma/\delta g_{\la\mu}(x) + \dots$,
where the dots stand for all the terms after $\nabla_\la \braket{T^{\la\mu}}$,
which do not contribute to the on-shell anomalies at lowest order.\footnote{Note that if the theory involves fermions, two additional contributions $\xi^\nu\omega_{\mu\hat\imath\hat\jmath}\braket{T^{[\hat\imath\hat\jmath]}}$ (where $\omega$ is the spin connection) and $(\nabla_\mu\xi_\nu)\braket{T^{[\hat\imath\hat\jmath]}}$ enter the diffeomorphism anomaly. However, both can be absorbed in the Lorentz anomaly (see \sec{sec:LorentzAnomaly}).
}
Similarly to the gauge current, assumption~\ref{Covariant} allows us to supplement the EMT by a BZ polynomial~\cite{Bardeen:1984pm,Bertlmann:1996xk} to obtain the covariant form of the anomalies
\begin{subequations} \begin{align}
\label{eq:DiffeoAnomalyPureGauge3pt}
(2\pi)^4 & \delta^4(p_1+p_2+p_3)\,{\cal A}^{ab\mu}_{AA}(1^{h_1}\!,2^{h_2}\!,3) \coloneqq \\*
{-}2i & \varepsilon_{1\nu} \varepsilon_{2\rho} p_{3\la} \frac{\delta\Gamma[g,A,0]}{\delta A^a_\nu(p_1) \delta A^b_\rho(p_2) \delta g_{\la\mu}(p_3)} \bigg|_{g=\eta,\,A=0,\,p_i^2=0}^\text{cov}\!, \nn \\
\label{eq:DiffeoAnomalyMixed3pt}
(2\pi)^4 & \delta^4(p_1+p_2+p_3)\,{\cal A}^{b\mu}_{gA}(1^{h_1}\!,2^{h_2}\!,3) \coloneqq \\*
{-}2i\kappa\,& \varepsilon_{1\nu\rho} \varepsilon_{2\sigma} p_{3\la} \frac{\delta\Gamma[g,A,0]}{\delta g_{\nu\rho}(p_1) \delta A^b_\sigma(p_2) \delta g_{\la\mu}(p_3)} \bigg|_{g=\eta,\,A=0,\,p_i^2=0}^\text{cov}\!, \nn \\
\label{eq:DiffeoAnomalyPureGrav3pt}
(2\pi)^4 & \delta^4(p_1+p_2+p_3)\,{\cal A}^\mu_{gg}(1^{h_1}\!,2^{h_2}\!,3) \coloneqq \\*
{-}2i\kappa^2 & \varepsilon_{1\nu\rho} \varepsilon_{2\sigma\tau} p_{3\la} \frac{\delta\Gamma[g,0,0]}{\delta g_{\nu\rho}(p_1) \delta g_{\sigma\tau}(p_2) \delta g_{\la\mu}(p_3)} \bigg|_{g=\eta,\,p_i^2=0}^\text{cov}\!. \nn
\end{align} \label{eq:DiffeoAnomalies3pt}%
\end{subequations}
Note that, contrary to the previous dimension-2 anomalies, these form factors have dimension 3, since $[\xi^\mu]=-1$.

We bootstrap these anomalies as bispinors of the most general form
${\cal A}^{\dot\gamma\gamma}\coloneqq\sigma^{\dot\gamma\gamma}_\mu {\cal A}^\mu={\cal M}_1 |1]^{\dot\gamma} \bra{1}^\gamma + {\cal M}_2 |2]^{\dot\gamma} \bra{2}^\gamma$.
Indeed, on the 3-point kinematics the cross-terms $|1]\bra{2}$ and $|2]\bra{1}$ are both proportional to either $|1]\bra{1}$ or $|2]\bra{2}$.
The scalar anomaly coefficients ${\cal M}_{1,2}$ may then be bootstrapped by \eqn{eq:General3pt} again assuming zero helicity for the form-factor leg 3.
So we use the polynomial assumption~\ref{Polynomial} and find in the pure-gauge sector~\eqref{eq:DiffeoAnomalyPureGauge3pt}
\begin{subequations} \begin{align}
\label{eq:DiffeoAnomalyGaugeBrakets1}
{\cal A}^{ab\mu}_{AA}(1^-\!,2^-\!,3) &
 = \delta^{ab} (C_1 p_1^\mu + C_2 p_2^\mu) \braket{12}^2 , \\*
\label{eq:DiffeoAnomalyGaugeBrakets2}
{\cal A}^{ab\mu}_{AA}(1^+\!,2^+\!,3) & = \delta^{ab} (\tilde{C}_1 p_1^\mu + \tilde{C}_2 p_2^\mu) [12]^2 , \\*
\label{eq:DiffeoAnomalyGaugeBrakets3}
{\cal A}^{ab\mu}_{AA}(1^+\!,2^-\!,3) & = {\cal A}^{ab\mu}_{AA}(1^-\!,2^+\!,3) = 0 .
\end{align} \label{eq:DiffeoAnomalyGaugeBrakets}%
\end{subequations}
Note that Bose symmetry implies $C_1 = C_2$, $\tilde{C}_1 = \tilde{C}_2$, which means that ${\cal A}^{ab\mu}_{AA} \propto p_1^\mu+p_2^\mu = -p_3^\mu$,
and the same happens in the pure-gravity sector~\eqref{eq:DiffeoAnomalyPureGrav3pt} but with doubled helicities.
However, the on-shell and transversality constraints on both metric perturbations $h_{\mu\nu}$ and $h'_{\mu\nu}=h_{\mu\nu}+(\partial_\mu\xi_\nu+\partial_\nu\xi_\mu)/\kappa$ imply transversality of $\xi^\mu$ as well, \ie the residual coordinate transformations satisfy $p_3 \cdot \xi(p_3)=0$.
We conclude that any theory involving gravity and gauge fields is free from diffeomorphism anomalies, at least on shell:
\begin{equation}
\xi_\mu {\cal A}^{ab\mu}_{AA} = \xi_\mu {\cal A}^\mu_{gg} = 0 .
\end{equation}
This agrees with the absence of diffeomorphism anomalies in pure gravity in four dimensions~\cite{Alvarez-Gaume:1983ihn}, and the fact that gauge theories do not violate general coordinate invariance in the absence of gravity.

In the mixed gauge-gravitational sector~\eqref{eq:DiffeoAnomalyMixed3pt}, we can use the polynomial assumption~\ref{Polynomial} to directly set to zero the mixed-helicity configurations: ${\cal A}^{b\mu}_{gA}(1^{+2}\!,2^{-1}\!,3) = {\cal A}^{b\mu}_{gA}(1^{-2}\!,2^{+1}\!,3) = 0$.
Indeed, all the candidate spinor expressions~\eqref{eq:General3pt} here have multiple kinematic denominators, as was the case for \eqn{eq:DiffeoAnomalyGaugeBrakets3}.
The other two cases, however, require a more careful treatment.
A naive application of \eqn{eq:General3pt} gives a single denominator, such as ${\cal A}^{b\mu}_{gA}(1^{-2}\!,2^{-1}\!,3) \propto p_1^\mu \braket{12}^3\braket{31}/\braket{23}$, where we have discarded the term along~$p_3^\mu$.
But such a denominator may be absorbed by a judicious choice of the basis vector:
\begin{subequations}
\begin{align}
\label{eq:DiffeoAnomalyMixedBrakets1}
{\cal A}^{b\mu}_{gA}(1^{-2}\!,2^{-1}\!,3) &
 = \kappa \tr t^b_r\,C_- \bra{1}\sigma^\mu|2] \braket{12}^3 , \\
\label{eq:DiffeoAnomalyMixedBrakets2}
{\cal A}^{b\mu}_{gA}(1^{+2}\!,2^{+1}\!,3) &
 = \kappa \tr t^b_r\,C_+ [1|\bar\sigma^\mu\ket{2} [12]^3 , 
\end{align} \label{eq:DiffeoAnomalyMixedBrakets}%
\end{subequations}
which is therefore not ruled out by~\ref{Polynomial}.

Here the gauge field must clearly be abelian and, by ${\cal C}$ invariance in the Dirac-fermion case, axial.
Moreover, ${\cal CPT}$ symmetry implies $C_+=-C_-^* \eqqcolon iQ'C$, whereas parity converts
${\cal P A}^\mu_{g\text{A}}(1^{-2}\!,2^{-1}\!,3) = -{\cal A}_\mu^{g\text{A}}(1^{+2}\!,2^{+1}\!,3)$.
The on-shell application of parity, however, gives
\begin{equation}
{\cal P} \big\{ i\kappa Q' C^* \bra{1}\sigma^\mu|2] \braket{12}^3 \big\}
 = -i\kappa Q' C^* [1|\bar\sigma_\mu\ket{2}  [12]^3 .
\label{eq:DiffeoAnomalyParity}
\end{equation}
Comparing this with (minus) \eqn{eq:DiffeoAnomalyMixedBrakets2}, we find $C\in\mathbb{R}$.
We have thus constrained the on-shell anomalies~\eqref{eq:DiffeoAnomalyMixedBrakets} down to a single constant, and in view of \ref{OneLoop}, the diffeomorphism anomaly can be uplifted to
\begin{equation}\!\!
\delta^{\text{d}}_\xi{} \Gamma\big|^{\text{cov}}\!= -4\sqrt{2} C\!\!\int\!\!\dd^4x \sqrt{|g|} (\nabla^\la \xi^\mu) R_{\la\mu\nu\rho} \widetilde{F}_\text{A}^{\nu\rho}\!\!\!\!\sum_{\text{fermions }\psi}\!\!\!\!Q'_\psi ,
\label{eq:DiffeoVariationAnswer}
\end{equation}
in agreement with \cite{Nieh:1984vx,Larue:2023qxw}.
This expression may also be brought closer to \eqns{eq:GaugeVariationAnswer}{eq:GaugeGravVariationAnswer} by rewriting
$2F_\text{A} \sum_\varphi Q'_\varphi  = F_\text{R} \sum_{\tilde\eta}\!Q_{\tilde\eta} - F_\text{L} \sum_\chi\!Q_\chi$.
The diffeomorphism-anomaly cancellation condition is the same as that of the axial-gravitational anomaly~\eqref{eq:axialgravCancellation}.

\section{Lorentz anomaly}
\label{sec:LorentzAnomaly}

When considering fermions on curved spacetime, one is forced to choose local tangent frames and thus switch to tetrad fields $e^{\hat\imath}_\mu$ as the fundamental degrees of freedom, such that $e^{\hat\imath}_\mu e^{\hat\jmath}_\nu \eta_{\hat\imath\hat\jmath} = g_{\mu\nu}$. 
The new freedom in choosing such frames corresponds to local Lorentz transformations
\begin{equation}
\delta^\text{L}_\xi e^{\hat\imath}_\mu = \xi^{\hat\imath}{}_{\hat\jmath} e^{\hat\jmath}_\mu , \qquad \quad
\xi_{\hat\imath\hat\jmath} = -\xi_{\hat\jmath\hat\imath} .
\label{eq:LorentzTransformation}
\end{equation}
This must be a symmetry of the classical theory.
This is equivalent to having a symmetric EMT, and at the quantum level it is now defined as
$\sqrt{|g|} \braket{T^{\hat\imath\hat\jmath}} \coloneqq e^{\hat\imath\mu} \delta\Gamma/\delta e_{\hat\jmath}^\mu$.

At the quantum level, this symmetry may be anomalous and is expressed as
\begin{equation}
\delta^\text{L}_\xi\Gamma[e^{\hat\imath}_\mu,A^a_\mu,\varphi] =\!\int\!\dd^4x \bigg[{-}\sqrt{|g|}\,\xi_{\hat\imath\hat\jmath} \braket{T^{\hat\imath\hat\jmath}} + \frac{\delta\Gamma}{\delta\varphi} \delta^\text{L}_\xi\varphi \bigg] .
\end{equation}
Analogously to \eqn{eq:GaugeAnomaly}, the Lorentz-anomaly functional is
$\smash{{\cal A}^{[\hat\imath\hat\jmath]}(x)\coloneqq -\delta(\delta^\text{L}_\xi \Gamma)/\delta\xi_{\hat\imath\hat\jmath}(x) = e^{\la[\hat\imath} \delta\Gamma/\delta e_{\hat\jmath]}^\la(x)} + \dots$,
where the dots stand for the matter terms following $\braket{T^{[\hat\imath\hat\jmath]}}$.
By assumption~\ref{Covariant}, we look for covariant forms of the Lorentz anomaly, which must therefore be expressed directly in terms of the metric.
Hence the relevant 3-point on-shell anomaly form factors (of dimension 2) are\footnote{Contrary to the diffeomorphism anomalies~\eqref{eq:DiffeoAnomalies3pt}, the gauge parameter~$\xi_{\hat\imath\hat\jmath}$ must not be transverse even on shell, since on-shell tetrads are only transverse in the (co)vector index: $\smash{\partial^\mu e^{\hat\imath}_\mu = 0}$.
}
\begin{subequations} \begin{align}
\label{eq:LorentzAnomalyPureGauge3pt}
& (2\pi)^4 \delta^4(p_1+p_2+p_3)\,{\cal A}^{ab[\hat\imath\hat\jmath]}_{AA}(1^{h_1}\!,2^{h_2}\!,3) \coloneqq \\* &
\varepsilon_{1\mu} \varepsilon_{2\nu} e_0^{\la[\hat\imath}(p_3) \frac{\delta\Gamma[e,A,0]}{\delta A^a_\mu(p_1) \delta A^b_\nu(p_2) \delta e_{\hat\jmath]}^\la(p_3)} \bigg|_{e=e_0,\,A=0,\,p_i^2=0}^\text{cov}\!, \nn \\
\label{eq:LorentzAnomalyMixed3pt}
& (2\pi)^4 \delta^4(p_1+p_2+p_3)\,{\cal A}^{b[\hat\imath\hat\jmath]}_{gA}(1^{h_1}\!,2^{h_2}\!,3) \coloneqq \\* &
\kappa\,\varepsilon_{1\mu\nu} \varepsilon_{2\rho} e_0^{\la[\hat\imath}(p_3) \frac{\delta\Gamma[e,A,0]}{\delta g_{\mu\nu}(p_1) \delta A^b_\rho(p_2) \delta e_{\hat\jmath]}^\la(p_3)} \bigg|_{e=e_0,\,A=0,\,p_i^2=0}^\text{cov}\!, \nn \\
\label{eq:LorentzAnomalyPureGrav3pt}
& (2\pi)^4 \delta^4(p_1+p_2+p_3)\,{\cal A}^{[\hat\imath\hat\jmath]}_{gg}(1^{h_1}\!,2^{h_2}\!,3) \coloneqq \\* &
\kappa^2 \varepsilon_{1\mu\nu} \varepsilon_{2\rho\sigma} e_0^{\la[\hat\imath}(p_3) \frac{\delta\Gamma[e,0,0]}{\delta g_{\mu\nu}(p_1) \delta g_{\rho\sigma}(p_2) \delta e_{\hat\jmath]}^\la(p_3)} \bigg|_{e=e_0,\,p_i^2=0}^\text{cov}\!, \nn
\end{align} \label{eq:LorentzAnomalies3pt}%
\end{subequations}
where $e_0$ denote flat tetrads, such as $e_0{}^{\hat\imath}_\mu = \delta^{\hat\imath}_\mu$.

Due to the antisymmetry of these anomalies ${\cal A}^{[\hat\imath\hat\jmath]}$ in the flat Minkowski indices $\hat\imath$ and $\hat\jmath$, they may be reduced to a couple of left- and right-handed symmetric bispinors:
\begin{equation}
\bar\sigma_{\hat\imath}^{\dot\gamma\gamma} \bar\sigma_{\hat\jmath}^{\dot\delta\delta} {\cal A}^{[\hat\imath\hat\jmath]}
 = {\cal A}^{(\gamma\delta)} \epsilon^{\dot\gamma\dot\delta} + {\cal A}^{(\dot\gamma\dot\delta)} \epsilon^{\gamma\delta} ,
\label{eq:ALdecomp}
\end{equation}
as reviewed in more detail in \app{app:Multispinors}.
On shell, they may be decomposed in the basis of 3 momentum bispinors, such as $\bra{1}^\gamma \bra{1}^\delta$, $\bra{2}^\gamma \bra{2}^\delta$ and $\bra{1}^{(\gamma} \bra{2}^{\delta)}$ for the left-handed form factor, each absorbing one mass dimension and different helicities.
In the pure-gauge sector~\eqref{eq:LorentzAnomalyPureGauge3pt}, the MHV bootstrap~\eqref{eq:General3ptMHV} thus gives\footnote{In \eqns{eq:AALorentz}{eq:gALorentz}, the on-shell relation $|1] = -|2] \braket{23}/\braket{13}$ reduces the ansatz to a single term proportional to a dimensionful constant~$C_4$, indicating a potential higher-dimensional operator.
}
\begin{subequations} \begin{align}
\label{eq:AALorentz1}
 & {\cal A}^{ab(\gamma\delta)}_{AA}(1^-\!,2^-\!,3) = \delta^{ab} \Big\{ C_1 \tfrac{\braket{12}\braket{23}}{\braket{13}} \bra{1}^\gamma \bra{1}^\delta \\* & \qquad\!\!\qquad
 + C_2 \tfrac{\braket{21}\braket{13}}{\braket{23}} \bra{2}^\gamma \bra{2}^\delta 
 + C_3 \braket{12} \bra{1}^{(\gamma} \bra{2}^{\delta)} \Big\} , \nn \\*
\label{eq:AALorentz2}
 & {\cal A}^{ab(\dot\gamma\dot\delta)}_{AA}(1^-\!,2^-\!,3) = C_4\,\delta^{ab} \braket{12}^3 |1]^{(\dot\gamma} |2]^{\dot\delta)} .
\end{align} \label{eq:AALorentz}%
\end{subequations}
All other helicity configurations result in at least one denominator in front of all three basis bispinors and are thus ruled out by the polynomial assumption~\ref{Polynomial}, which also sets $C_1=C_2=0$.
Moreover, Bose symmetry $1 \leftrightarrow 2$ implies $C_3=C_4=0$.
Since the same argument holds on the $\smash{\overline{\text{MHV}}}$ branch,
we find that the pure-gauge Lorentz anomaly $\smash{{\cal A}^{ab[\hat\imath\hat\jmath]}_{AA}}$ vanishes.
The same chain of reasoning, unsurprisingly, rules out a Lorentz anomaly~$\smash{{\cal A}^{[\hat\imath\hat\jmath]}_{gg}}$ in the pure-gravity sector.

The cross-sector~\eqref{eq:LorentzAnomalyMixed3pt} on the MHV branch is
\begin{subequations} \begin{align}
\label{eq:gALorentz1}
{\cal A}^{b(\gamma\delta)}_{gA}(1^{-2}\!,2^{-1}\!,3) &
 = \kappa \tr t_r^b \Big\{ C_1 \braket{12}^2 \bra{1}^\gamma \bra{1}^\delta \\*
 +\,C_2 \tfrac{\braket{12}^2\braket{31}^2}{\braket{23}^2\!} & \bra{2}^\gamma \bra{2}^\delta
 + C_3 \tfrac{\braket{12}^2\braket{31}}{\braket{23}} \bra{1}^{(\gamma} \bra{2}^{\delta)} \Big\} , \nn \\
\label{eq:gALorentz2}
{\cal A}^{b(\dot\gamma\dot\delta)}_{gA}(1^{-2}\!,2^{-1}\!,3) &
 = \kappa\,C_4 \tr t_r^b\,\braket{12}^4 |2]^{\dot\gamma} |2]^{\dot\delta} ,
\end{align} \label{eq:gALorentz}%
\end{subequations}
where $C_2=C_3=0$ by \ref{Polynomial}, and similarly for the $\smash{\overline{\text{MHV}}}$ branch.
The color factor $\tr t_r^a$ and the \ref{CParity} assumption again restrict these anomalies to abelian axial gauge fields.
The right-handed MHV term~\eqref{eq:gALorentz2} requires $[C_4]=-2$ and is actually a so-called trivial anomaly, which can be renormalized away~\cite{Bertlmann:1996xk}, as we show in \app{app:Lorentz} by presenting an explicit local counterterm.
Therefore, we can focus on the dimension-2 MHV and $\smash{\overline{\text{MHV}}}$ terms $\kappa Q' C_- \braket{12}^2 \bra{1} \bra{1}$ and $\kappa Q'C_+ [12]^2 |1] |1]$.
In view of the definition~\eqref{eq:LorentzAnomalyMixed3pt}, they must be related via\footnote{The second sign in \eqn{eq:LorentzAnomalyParity} comes from the ${\cal P}$-oddness of $\delta/\delta A_\text{A}^\mu$ and the spinor relation~\eqref{eq:BispinorParity1}.
}
\begin{equation}
{\cal CPT A}_{g\text{A}}^{(\gamma\delta)} = -{\cal A}^{(\dot\gamma\dot\delta)}_{g\text{A}}\!, \qquad
{\cal P A}_{g\text{A}}^{(\gamma\delta)} = -\epsilon_{\dot\gamma\dot\varepsilon} \epsilon_{\dot\delta\dot\varphi} {\cal A}^{(\dot\varepsilon\dot\varphi)}_{g\text{A}}\!,
\label{eq:LorentzAnomalyParity}
\end{equation}
which give $C_-^* = -C_+$ and $C_-=-C_+$, respectively.
Hence we have one real unfixed coefficient $C\coloneqq\pm C_\pm$.
By \ref{OneLoop} additivity, we uplift the Lorentz anomaly to
\begin{equation}
\delta^{\text{L}}_\xi{} \Gamma\big|^{\text{cov}}\!= -\sqrt{2}C\!\!\int\!\!\dd^4x \sqrt{|g|}\,\xi_{\hat\imath\hat\jmath} R^{\hat\imath\hat\jmath}{}_{\mu\nu} \widetilde{F}_\text{A}^{\mu\nu}\!\!\!\!\sum_{\text{fermions }\psi}\!\!\!\!Q'_\psi ,
\label{eq:LorentzVariationAnswer}
\end{equation}
which is in agreement with the literature~\cite{Leutwyler:1984nd,Leutwyler:1985ar,Nieh:1984vx,Caneschi:1985an,Yajima:1986bi,Larue:2023qxw}.
The Lorentz-anomaly cancellation condition is the same as that of the mixed axial-gravitational anomaly~\eqref{eq:axialgravCancellation}.

\section{Weyl anomaly}
\label{sec:WeylAnomaly}

So far we have focused on local-symmetry anomalies, though the gauge anomaly~\eqref{eq:GaugeVariationAnswer} may of course be reduced to the global chiral anomaly by switching off the axial field and restricting to constant gauge parameters $\xi_\text{A}=[g_\text{R}\xi_\text{R}-g_\text{L}\xi_\text{L}]/g'$.
To demonstrate that our approach is also applicable to purely global anomalies, we now bootstrap the Weyl (trace) anomaly.
The Weyl transformation rescales the fields (but not the coordinates and gauge connections) and infinitesimally looks like
\begin{equation}
\delta^\text{W}_\xi\!g_{\mu\nu} = 2\xi g_{\mu\nu} , \qquad
\delta^\text{W}_\xi\!A^a_\mu = 0 , \qquad
\delta^\text{W}_\xi\!\varphi = \xi\Delta_\varphi \varphi ,
\label{eq:WeylTransformation}
\end{equation}
where $\Delta_\varphi$ is a constant scaling dimension that depends on the nature of~$\varphi$.
A theory is classically Weyl-invariant if and only if the EMT is traceless $T^\mu{}_\mu=0$.\footnote{Since the EMT trace is not sensitive to its antisymmetric part, we take the metric as the fundamental field instead of the tetrad. 
}
Although it would not put the consistency of the theory in jeopardy, this symmetry could be anomalous:
\begin{equation}
\delta^\text{W}_\xi\Gamma[g,A,\varphi]
 =\!\int\!\dd^4x\,\xi\bigg[{-}\sqrt{|g|} \braket{T^\mu{}_\mu} + \Delta_\varphi \varphi \frac{\delta\Gamma}{\delta\varphi}  \bigg] .
\end{equation}
Here we allowed the gauge parameter to be local, which is standard practice when deriving Noether currents.
The Weyl-anomaly functional
${\cal A}(x)\coloneqq -\delta(\delta^\text{W}_\xi \Gamma)/\delta\xi(x) = -2g_{\la\mu} \delta\Gamma/\delta g_{\la\mu}(x) + \dots$
implies definitions for the on-shell anomaly amplitudes that are identical to \eqn{eq:DiffeoAnomalies3pt} except for the replacement $ip_{3\la} \to \eta_{\la\mu}$.

The 3-point Weyl-anomaly amplitudes are Lorentz scalars of dimension~2, so the kinematic bootstrap~\eqref{eq:General3pt} is identical to the gauge-anomaly cases.
This is enough to immediately rule out the mixed gauge-gravitational anomalies~${\cal A}_{gA}^b=0$.\footnote{Namely, the mixed gauge-gravitational Weyl anomalies correspond to nonpolynomial expressions, such as $\smash{{\cal A}_{gA}^b(1^{-2}\!,2^{-1}\!,3)} \propto \smash{\kappa \braket{12}^3\braket{31}/\braket{23}}$, which are ruled out by our assumption~\ref{Polynomial}.
}
In the pure-gauge sector, we get
\be
\label{eq:WeylAnomalyGaugeBrakets}
{\cal A}_{AA}^{ab}(1^\mp\!,2^\mp\!,3) = g^2 \delta^{ab} \times \Big\{\!
\begin{smallmatrix}
C^- \braket{12}^2 & \text{ for } (-,-) , \\
C^+ [12]^2 & \text{ for } (+,+) .
\end{smallmatrix}
\ee
The nonkinematic considerations are now straightforward: the parity relation between the helicity configurations, which holds in the vector-like Dirac-fermion case, gives $C^+=C^- \eqqcolon C_1$, and ${\cal CPT}$ invariance implies $C_1 \in \mathbb{R}$.
Similarly, the purely gravitational anomaly is
\be
\label{eq:WeylAnomalyGravBrakets}
{\cal A}_{gg}(1^\mp\!,2^\mp\!,3) = \kappa^2 C_2 \times \Big\{\!
\begin{smallmatrix}
\braket{12}^4 & \text{ for } (-,-) , \\
[12]^4 & \text{ for } (+,+) ,
\end{smallmatrix} \quad C_2 \in \mathbb{R} .
\ee
Unlike the previous cases, we do not obtain any additional constraints on which particles in the loop result in this anomaly, \ie any (massless) particle may contribute to it, so the gauge field in \eqn{eq:WeylAnomalyGaugeBrakets} is a generic one.

Now if we consider a single Dirac fermion, which is classically gauged vectorially and axially, we can deduce $C_\text{L}^\pm = C_\text{R}^\pm$ from its ${\cal C}$ invariance ${\cal C A}_\text{LL}^{ab}= \smash{\tfrac{g_\text{L}^2}{g_\text{R}^2}} {\cal A}_\text{RR}^{ab}$.
Using the assumption~\ref{Covariant}, \eqn{eq:FFonshell}, its gravitational analogue, and \ref{OneLoop} additivity, we can now uplift the Weyl anomaly off shell:
\begin{widetext}
\begin{equation}
\label{eq:WeylVariationAnswer}
\delta^{\text{W}}_\xi \Gamma = \xi \!\int\!\dd^4x \sqrt{|g|}\,\bigg\{
   \frac{C_1}{2} \Big[\!\sum_\chi g_\chi^2 \tr\!{}_{r_\chi} (F^\text{L}_{\mu\nu} F_\text{L}^{\mu\nu}) + \sum_{\tilde\eta} g_{\tilde\eta}^2 \tr\!{}_{r_{\tilde\eta}} (F^\text{R}_{\mu\nu} F_\text{R}^{\mu\nu})\Big]
 + \sum_\phi C_1^{\phi} g_\phi^2 \tr\!{}_{r_\phi} (F_{\mu\nu} F^{\mu\nu}) 
 - R_{\la\mu\nu\rho} R^{\la\mu\nu\rho} \sum_{\varphi'}\!C_2^{\varphi'}\!
   \bigg\} .
\end{equation}
\end{widetext}
Here $\varphi'$ goes over all massless particles, and $\phi$ stands for all massless particles except the Weyl fermions.
\Eqn{eq:WeylVariationAnswer} is of course insensitive to other terms vanishing on shell, such as $R^2, R_{\mu\nu} R^{\mu\nu}$ and $D^2 R$.

It is remarkable that this simple bootstrap is able to predict that the Weyl anomaly does not involve the Pontryagin densities $F\cdot\widetilde{F}$ and $R\cdot\widetilde{R}$.
This fact is implied simply by $C^+=C^-$, which we derived even before employing ${\cal CPT}$ invariance.
The question whether Weyl fermions could produce such densities has been a subject of debate over the past decade~\cite{Bonora:2014qla,Bonora:2015nqa,Bastianelli:2016nuf,Bonora:2017gzz,Bonora:2018obr,Bastianelli:2019zrq,Frob:2019dgf,Abdallah:2021eii,Bonora:2022izj,Abdallah:2023cdw,Larue:2023qxw,Larue:2023tmu,Alvarez:2025mql,Anero:2026pub}.
Although there is a model-independent resolution in the literature~\cite{Larue:2023qxw}, we have just provided proof that the Weyl anomaly cannot be a Pontryagin density, which is both model- and regularization-free, and is based on the minimal assumption~\ref{CParity} that ${\cal P}$ invariance does not break at the quantum level in case it is there classically.
Note that this assumption is consistent with the literature, as Pontryagin densities have only been claimed to appear in the scenarios where ${\cal C}$ and ${\cal P}$ are violated already at the classical level, such as for a single Weyl fermion.
Moreover, our purely four-dimensional argument implies that any independent loop-integration calculation performed using a regularization scheme, which is consistent with our assumptions, should not give Pontryagin densities either.

\section{Discussion and outlook}
\label{sec:outro}

In this paper, we have bootstrapped the chiral, mixed axial-gravitational, diffeomorphism, Lorentz, and Weyl anomalies using the massless spinor-helicity formalism~\cite{Berends:1981rb,DeCausmaecker:1981jtq,Gunion:1985vca,Kleiss:1985yh,Xu:1986xb,Gastmans:1990xh}, up to real constant prefactors and terms that vanish on shell.
Although the formalism involves several conventional, interrelated phase factors, the final results are independent of these choices, as explained in \app{app:PT}.
Our findings suggest that for anomalies the traditional distinction between kinematics and dynamics is less sharp than is often assumed.
Just as scattering amplitudes can be characterized by symmetry, factorization, and consistency, quantum anomalies also admit a similar bootstrap description.
From this perspective, anomalies are not primarily consequences of a particular regularization or calculation scheme, but of the consistency conditions that any QFT must satisfy.

Several questions remain open.
Firstly, it would be desirable to derive the one-loop exactness~\cite{Adler:1969er} of the anomalies (excluding the Weyl anomaly) directly within the on-shell framework, without invoking regularization.
Moreover, we have not considered matter particles on the external legs within our bootstrap, which, strictly speaking, could provide additional information if not immediately ruled out.
Furthermore, one may ask what happens if the 3-point anomaly amplitudes are used inside on-shell recursion in a broader scattering-amplitude context.
On the one hand, this could simply reconstruct the higher-point contributions that we automatically included by our covariance assumption~\ref{Covariant}.
On the other hand, this might help probe the terms that we missed in the Weyl anomaly due to working at the 3-point level.
Finally, since global anomalies do not invalidate the original field theory, such an on-shell recursion may be combined with other on-shell methods to bootstrap scattering amplitudes for Goldstone bosons, which can couple to anomalous currents, such as axions~\cite{Peccei:1977hh,Peccei:1977ur,Weinberg:1977ma,Wilczek:1977pj}.

\section*{Acknowledgements}
We thank Sixun Du, Alexey Koshelev, Panagiotis Marinellis, Alex Pomarol, and J\'er\'emie Quevillon for helpful conversations.
HK is supported by the National Natural Science Foundation of China via NSFC grant No. 12447147.
HK, RL, and AO are supported by the Science and Technology Commission of the Shanghai
Municipality via STCSM grant No. 24ZR1450600.

\appendix
\section{Effective action}
\label{app:Effective}

In this paper, we rely on the effective action $\Gamma[\varphi]$ (written in terms of abstract fields $\varphi_i(x)$),
which is defined by the functional Legendre transform
\be
\Gamma[\varphi]=W[J] -\!\int\!\dd^4x\,J(x)\!\cdot\!\varphi(x) \Big|_{\varphi=\delta W/\delta J}
\label{eq:LegendreTransform}
\ee
of the generating functional $W[J]\coloneqq-i\log\big(Z[J]/Z[0]\big)$ for connected Green's functions,
\ie
\be
e^{iW[J]} = \frac{\int\!{\cal D}\varphi\,e^{i(S[\varphi]+\int\!\dd^4x J(x)\cdot\varphi(x))}}{\int\!{\cal D}\varphi\,e^{iS[\varphi]}} .
\ee
In particular, the field-argument substitution in the definition~\eqref{eq:LegendreTransform} $\delta W[J]/\delta J(x) = \braket{\Omega|\varphi(x)|\Omega}_J$ is by the mean field in the presence of the classical background current~$J$.

We normalize scalars $\phi(x)$, gauge fields~$A_\mu(x)$ and metric perturbations~$h_{\mu\nu}(x)\coloneqq[g_{\mu\nu}(x)-\eta_{\mu\nu}]/\kappa$ all to have dimension~1, in particular
\be
\frac{\delta}{\delta g_{\mu\nu}(x)} = \frac{1}{\kappa} \frac{\delta}{\delta h_{\mu\nu}(x)} ,
\label{eq:MetricDerivatives}
\ee
where $\kappa\coloneqq\sqrt{32\pi G_\text{Newton}}$ accounts for the mass dimension of the gravitational field.
The relevant Fourier transforms have less familiar mass dimensions
\begin{subequations}
\beal \relax
[\phi(p)] = [A_\mu(p)] = [h_{\mu\nu}(p)] & = -3 , \\
[F_{\mu\nu}(p)] = [R_{\la\mu\,\nu\rho}(p)] & = -2 ,
\label{eq:MassDimensions1}
\eeal
where the Riemann tensor's dimension includes $[\kappa]=-1$.
The functional derivatives, however, undo dimensionful integrals ($\dd^4x$ or $\dd^4p$), so their mass dimensions are
\beal
\bigg[\frac{\delta}{\delta \varphi(x)}\bigg] & = \bigg[\frac{\delta}{\delta A_\mu(x)}\bigg]
 = \bigg[\frac{\delta}{\delta h_{\mu\nu}(x)}\bigg] = 3 , \\
\bigg[\frac{\delta}{\delta \varphi(p)}\bigg] & = \bigg[\frac{\delta}{\delta A_\mu(p)}\bigg]
 = \bigg[\frac{\delta}{\delta h_{\mu\nu}(p)}\bigg] = -1 .
\label{eq:MassDimensions2}
\eeal
\label{eq:MassDimensions}%
\end{subequations}
In the main text, we favor the derivatives in the metric, so we expose gravitational constants by hand according to \eqn{eq:MetricDerivatives}, see \eg \eqn{eq:GaugeGravAnomaly3pt}.

\section{Multispinors}
\label{app:Multispinors}

The on-shell helicity spinors are defined by the equations recalled in 
\foot{foot:SpinorHelicity}
only up to ${\rm U}(1)$ transformations, which may be chosen independently for each momentum~$p$.
More generally, the Pauli matrices allow to trade any tensor index for a pair of ${\rm SL}(2,\mathbb{C})$ indices \cite{Penrose:1987uia}.
For antisymmetric tensors, the spinor map is particularly advantageous, as the 4-spinor decomposes neatly into the left- and right-handed pieces:
\be
F^{\dot\alpha\alpha\dot\beta\beta}
\coloneqq F^{\mu\nu} \bar\sigma_\mu^{\dot\alpha\alpha} \bar\sigma_\nu^{\dot\beta\beta}
 = \epsilon^{\dot\alpha\dot\beta} F^{\alpha\beta} + \bar{F}^{\dot\alpha\dot\beta} \epsilon^{\alpha\beta} .
\ee
Here the bispinors $F^{\alpha\beta}$ and $\bar{F}^{\dot\alpha\dot\beta}$ are automatically symmetric and can be obtained by contraction with the left- and right-handed ${\rm SL}(2,\mathbb{C})$ Lorentz generators:
\be\!\!\!
F^{\alpha\beta}\!= -iF_{\mu\nu} \epsilon^{\alpha\gamma} \sigma^{\mu\nu,}{}_\gamma{}^\beta\!, \quad
\bar{F}^{\dot\alpha\dot\beta}\!= -iF_{\mu\nu} \bar\sigma^{\mu\nu,\dot\alpha}{}_{\dot\gamma} \epsilon^{\dot\gamma\dot\beta}\!,\!
\ee
where the generators are explicitly
\be
\sigma^{\mu\nu,}{}_\alpha{}^\beta \coloneqq \frac{i}{2}
   \sigma^{[\mu}_{\alpha \dot{\gamma}} \bar{\sigma}^{\nu],\dot{\gamma} \beta} , \quad
\bar{\sigma}^{\mu\nu,\dot{\alpha}}{}_{\dot{\beta}} \coloneqq \frac{i}{2}
   \bar{\sigma}^{[\mu,\dot{\alpha} \gamma} \sigma^{\nu]}_{\gamma \dot{\beta}} .
\label{eq:LorentzGenerators}
\ee
In view of the on-shell action of parity (see \tab{tab:CPT}), we also need to keep in mind that signs in the decomposition of the dual 4-spinor
\be
F_{\alpha\dot\alpha\beta\dot\beta}
\coloneqq F_{\mu\nu} \sigma^\mu_{\alpha\dot\alpha} \sigma^\nu_{\beta\dot\beta}
 = F_{\alpha\beta} \epsilon_{\dot\alpha\dot\beta} + \epsilon_{\alpha\beta} \bar{F}_{\dot\alpha\dot\beta}
\ee
are consistent with
\be
F_{\alpha\beta} = \epsilon_{\alpha\gamma} F^{\gamma\delta} \epsilon_{\delta\beta} , \quad
\bar{F}_{\dot\alpha\dot\beta} = \epsilon_{\dot\alpha\dot\gamma} \bar{F}^{\dot\gamma\dot\delta} \epsilon_{\dot\delta\dot\beta} .
\label{eq:BispinorConsistency}
\ee
However, we impose that the Levi-Civita tensors obey $\epsilon_{\alpha\beta} \epsilon^{\beta\gamma} = \delta_\alpha^\gamma$, which means that they differ by overall signs: $\epsilon^{\alpha\beta}=-\epsilon_{\alpha\beta}=\epsilon^{\dot\alpha\dot\beta}=-\epsilon_{\dot\alpha\dot\beta}$.
Therefore, comparing how parity acts on a ${\cal P}$-even or -odd antisymmetric tensor
\beal
{\cal P} F^{\dot\alpha\beta\dot\gamma\delta} &
 = {\cal P}  F^{\mu\nu} \bar\sigma_\mu^{\dot\alpha\beta} \bar\sigma_\nu^{\dot\gamma\delta} = \pm F_{\mu\nu} \bar\sigma_\mu^{\dot\alpha\beta} \bar\sigma_\nu^{\dot\gamma\delta} \\ &
 = \pm F_{\mu\nu} \sigma^\mu_{\alpha\dot\beta} \sigma^\nu_{\gamma\dot\delta}
 = \pm F_{\alpha\dot\beta\gamma\dot\delta}
\eeal
with \eqn{eq:BispinorConsistency}, we finally have
\begin{subequations}
\begin{align}
\label{eq:BispinorParity1}
{\cal P} F^{\beta\delta} & = \mp \bar{F}_{\dot\beta\dot\delta} = \pm \epsilon_{\dot\beta\dot\alpha} \epsilon_{\dot\delta\dot\gamma} \bar{F}^{\dot\alpha\dot\gamma} , \\
\label{eq:BispinorParity2}
{\cal P} \bar{F}^{\dot\alpha\dot\gamma} & = \mp F_{\alpha\gamma} = \pm \epsilon_{\alpha\beta} \epsilon_{\gamma\delta} F^{\beta\delta} .
\end{align} \label{eq:BispinorParity}%
\end{subequations}

\section{Discrete Lorentz transformations}
\label{app:PT}

Here we expand on the on-shell representations of the discrete transformations ${\cal C}$, ${\cal P}$, and ${\cal T}$, which we summarized in \tab{tab:CPT} and heavily relied upon in our bootstrap.

The second and third rows describe the transformations of momentum spinors, which are the most basic kinematic building blocks.
As such, their properties directly imply those for momenta and polarizations vectors (fourth and fifth rows, respectively) via the equivalent inverse spinor maps $p^\mu = \bra{p}\sigma^\mu|p]/2 = [p|\bar\sigma^\mu\ket{p}/2$ and the definitions~\eqref{eq:PolVectors}.
For instance, we have
\begin{subequations}
\begin{equation} \begin{aligned}
{\cal P} p^\mu = \smash{\frac{1}{2}} & \smash{\big\{{\cal P}\bra{p}^\alpha = -[p|_{\dot\alpha}\big\}} \\ \times &
\smash{\big\{\sigma_{\alpha\dot\beta}^\mu = \bar\sigma_\mu^{\dot\alpha\beta}\big\}} \\ \times &
\smash{\big\{{\cal P}|p]^{\dot\beta} = -\ket{p}_\beta\big\} = +\frac{1}{2}[p|\bar\sigma_\mu\ket{p}}
 = p_\mu ,
\label{eq:SampleP}
\end{aligned} \end{equation}
and
\begin{equation} \begin{aligned}\!
{\cal T} \varepsilon_{p,+}^\mu =\,& \smash{\big\{{\cal T}\bra{q}^\alpha = -\ket{q}_\alpha\big\}} \\ \times &
\smash{\big\{\big(\sigma_{\alpha\dot\beta}^\mu\big)^* = \sigma_{\beta\dot\alpha}^\mu = \bar\sigma_\mu^{\dot\beta\alpha}\big\}} \\ \times &
\smash{\big\{{\cal T}\,|p]^{\dot\beta} = -[p|_{\dot\beta}\big\}} \\ /&
\smash{\big(\sqrt{2}\big\{{\cal T}\bra{q}^\gamma = -\ket{q}_\gamma\big\}} \\ &\times\!
\smash{\big\{{\cal T}\ket{p}_\gamma = \bra{p}^\gamma\big\}\big) = -\frac{[p|\bar\sigma_\mu\ket{q}}{\sqrt{2}\braket{pq}}}
 = \varepsilon^+_{p,\mu} ,\!
\label{eq:SampleT}
\end{aligned} \end{equation} \label{eq:SamplePT}%
\end{subequations}
where the antiunitarity of time reversal resulted in complex conjugation, as shown in the last row of \tab{tab:CPT}.
Note that we have used
\begin{equation}
{\cal P} \bra{p}^\alpha = \epsilon^{\alpha\beta} {\cal P} \ket{p}_\beta = \epsilon^{\dot\alpha\dot\beta} |p]^{\dot\beta} = -[p|_{\dot\alpha} ,
\label{eq:Pbra}
\end{equation}
as implied by \tab{tab:CPT}.
The above properties agree with the textbook notion of how momenta and polarizations should transform (see \eg \cite{Peskin:1995ev,Schwartz:2014sze}).

Let us therefore focus on the spinor transformation rules.
They are convention-dependent, and are chosen to mesh well with a number of other commonly used identities.
First of all, we assume that the conjugation property
\begin{equation}
(\bra{p}_\alpha)^*= \sgn(p^0)[p|_{\dot\alpha}
\label{eq:SpinorConjugation}
\end{equation}
(also largely conventional) holds for real momenta.
This fixes the transformations of the right-handed spinor~$[p|$ with respect to those of $\ket{p}$, and those for $\bra{p}$ and $|p]$ follow as per \eqn{eq:Pbra}.
The conjugation rule~\eqref{eq:SpinorConjugation} is consistent with the explicit spinor parametrizations from \eg \cite{Maitre:2007jq,Ochirov:2018uyq}.
The sign choice in ${\cal P}\ket{p}_\alpha = |p]^{\dot\alpha}$ is purely conventional.
Indeed, an inverse logic is allowed:
if we assume the textbook parity transformation of momentum, derived in \eqn{eq:SampleP}, then it turns out to connect the two default explicit spinor parametrizations of \cite{Ochirov:2018uyq} in terms of momentum components, which differ by an irrelevant U(1) phase, \ie
$\ket{{\cal P}p}_\alpha \sim |p]^{\dot\alpha}$.
In other words, these symmetries may switch between different little-group gauge choices, so it is important to transform all spinors at once.
Note that the conjugation property~\eqref{eq:SpinorConjugation} does not hold for complex momenta, which are required for a nontrivial three-point bootstrap.
We may still use the same algebraic conventions of \tab{tab:CPT} as for real momenta and work in the standard analytic-continuation paradigm as the three-point amplitude bootstrap~\cite{Benincasa:2007xk,McGady:2013sga}, which assumes that the physics in question can be described consistently across different spacetime signatures.
This is similar to how path-integral anomaly calculations~\cite{Bertlmann:1996xk,Fujikawa:2004cx} are generally carried out in Euclidean signature before being analytically continued to Lorentzian signature.

Moreover, although in this paper we do not deal with any external fermions, we have made sure that the spinor conventions of \tab{tab:CPT} are consistent with the textbook ${\cal C}$, ${\cal P}$, and ${\cal T}$ transformations of quantum fermion fields
\begin{equation}\!\!\!\!\begin{aligned}
\chi(x) & =\!\!\int\!\!\!\frac{\dd^3\bm{p}}{\!(2\pi)^3 2|\bm{p}|} \ket{p}
   \big[ e^{-ip \cdot x} c_-(\bm{p})\:\!{+}\;\!e^{ip \cdot x} d_+^\dagger(\bm{p}) \big]_{p^0=|\bm{p}|} , \\*
\tilde\eta(x) & =\!\!\int\!\!\!\frac{\dd^3\bm{p}}{\!(2\pi)^3 2|\bm{p}|} |p]
   \big[ e^{-ip \cdot x} c_+(\bm{p})\:\!{+}\;\!e^{ip \cdot x} d_-^\dagger(\bm{p}) \big]_{p^0=|\bm{p}|} .
\label{eq:WeylFermions}
\end{aligned}\!\!\!\!\!\end{equation}
Namely, if we assume that parity acts on the chiral-fermion annihilation operators as
\begin{subequations} \begin{align}
U_{\cal P}\,c_{\pm,j}(\bm{p})\,U_{\cal P}^{-1} & = \mp \varrho\,c_{\mp,j}(-\bm{p}) , \\*
U_{\cal P}\,d_{\pm,\bar\jmath}(\bm{p})\,U_{\cal P}^{-1} & = \pm \varrho^* d_{\mp,\bar\jmath}(-\bm{p}) ,
\end{align} \label{eq:ParityWeylParticles}%
\end{subequations}
(where $\varrho$ is an arbitrary constant phase factor)
then the fields transform as
\begin{subequations} \begin{align}
U_{\cal P}\,\chi_{\alpha j}(x)\,U_{\cal P}^{-1} & = \varrho\,\tilde\eta^{\dot\alpha}_j(x^0,-\bm{x}) , \\
U_{\cal P}\,\tilde\eta^{\dot\alpha}_j(x)\,U_{\cal P}^{-1} & = \varrho\,\chi_{\alpha j}(x^0,-\bm{x}) .
\end{align} \label{eq:ParityWeyl}%
\end{subequations}
If they can be combined into a Dirac field, then its parity transformation will take the familiar~\cite{Peskin:1995ev,Schwartz:2014sze} form
$U_{\cal P} \Psi_j(x) U_{\cal P}^{-1} = \varrho\,\gamma^0\Psi_j(x^0,-\bm{x})$,
where we use the Weyl basis
$\gamma^\mu=\smash{\big(\begin{smallmatrix} 0 & \sigma^\mu \\ \bar{\sigma}^\mu & 0 \end{smallmatrix}\!\big)}$
for the Dirac matrices.
Similarly, if the time reversal acts on the particle operators as
\begin{subequations} \begin{align}
U_{\cal T}\,c_{\pm,j}(\bm{p})\,U_{\cal T}^{-1} & = \theta\,c_{\pm,j}(-\bm{p}) , \\
U_{\cal T}\,d_{\pm,\bar\jmath}(\bm{p})\,U_{\cal T}^{-1} & = \theta^* d_{\pm,\bar\jmath}(-\bm{p}) ,
\end{align} \label{eq:TimeReversalWeylParticles}%
\end{subequations}
(where $\theta$ encodes another constant phase)
then its antiunitary action on the fields implies
\begin{subequations} \begin{align}
U_{\cal T}\,\chi_{\alpha j}(x)\,U_{\cal T}^{-1} & = \theta\,\chi^\alpha_j(x)(-x^0,\bm{x}) , \\
U_{\cal T}\,\tilde\eta^{\dot\alpha}_j(x)\,U_{\cal T}^{-1} & = -\theta\,\tilde\eta_{\dot\alpha j}(-x^0,\bm{x}) ,
\end{align} \label{eq:TimeReversalWeyl}%
\end{subequations}
or the familiar
$U_{\cal T} \Psi_j(x) U_{\cal T}^{-1} = \theta\,\gamma^1\gamma^3\overline\Psi_{\bar\jmath}(-x^0,\bm{x})$
in the Dirac-spinor language.
In the derivation of \eqn{eq:TimeReversalWeyl}, one effectively has to combine the parity transformation with complex conjugation~\eqref{eq:SpinorConjugation},
\begin{subequations} \begin{align}
\ket{p}_\alpha ~\to~ & U_{\cal T} \big\{\ket{{\cal P}p}_\alpha {\sim}\,|p]^{\dot\alpha}\big\}U_{\cal T}^{-1} = \bra{p}^\alpha , \\
|p]^{\dot\alpha} ~\to~ & U_{\cal T} \big\{|{\cal P}p]^{\dot\alpha} {\sim}\,{-}\ket{p}_\alpha\big\}U_{\cal T}^{-1} = -[p|_{\dot\alpha} ,
\end{align} \label{eq:TimeReversalSpinors}%
\end{subequations}
which gives the ${\cal T}$ column of \tab{tab:CPT}.
Finally, the fermionic-field properties~\eqref{eq:ChargeConjugationDirac} and~\eqref{eq:ChargeConjugationWeyl} under charge conjugation are consistent with
\begin{subequations} \begin{align}
U_{\cal C}\,c_{\pm,j}(\bm{p})\,U_{\cal C}^{-1} & = \varkappa\,d_{\pm,\bar\jmath}(\bm{p}) , \\*
U_{\cal C}\,d_{\pm,\bar\jmath}(\bm{p})\,U_{\cal C}^{-1} & = \varkappa^* c_{\pm,j}(\bm{p}) .
\end{align} \label{eq:ChargeConjugationWeylParticles}%
\end{subequations}
Note that when one deals with objects involving external matter, not considered here,
their transformation properties under ${\cal C}$, ${\cal P}$, and~${\cal T}$ may actually be sensitive to the above phase factors $\varkappa$, $\varrho$, and $\theta$, respectively.
A similar set of exercises shows the consistency of the polarization-vector properties in the fifth row of \tab{tab:CPT} with the textbook vector-field properties of \tab{tab:VALR}.

Importantly, the results of our bootstrap throughout this paper are insensitive to the precise conventions of \tab{tab:CPT}, as long as they are consistent internally and with \tab{tab:VALR}, which we have just argued here.
Indeed, the conventional spinor-helicity phases appear on both sides of all of our ${\cal C}$, ${\cal P}$, and ${\cal T}$ constraints.
On the one hand, we read off the discrete transformations of each anomaly amplitude from its definition, such as
\begin{align}
\label{eq:GaugeAnomaly3ptChiral}
(2\pi)^4 & \delta^4(p_1+p_2+p_3)\,{\cal A}_\text{LLL/RRR}^{abc}(1^{h_1}\!,2^{h_2}\!,3) \coloneqq \\*
{-}i\varepsilon_1^\la & \varepsilon_2^\mu p_3^\nu \frac{\delta^3 \Gamma[A_\text{L},A_\text{R},0]}{\delta A_\text{L/R}^{a\la}(p_1) \delta A_\text{L/R}^{b\mu}(p_2) \delta A_\text{L/R}^{c\nu}(p_3)} \bigg|_{A_\text{L/R}=0,\,p_i^2=0}^\text{cov} \nn
\end{align}
(which is the chiral generalization of \eqn{eq:GaugeAnomaly3ptOnShell})
by matching the field derivatives to their transformations in \tab{tab:VALR}, and the momenta and polarization vectors to their properties in \tab{tab:CPT}.
This gives us the expected result for a given transformation, such as \eqn{eq:GaugeAnomalyParity1} for ${\cal P}$.
On the other hand, we apply the spinor transformations in \tab{tab:CPT} to an on-shell kinematic ansatz, as in \eqn{eq:GaugeAnomalyParity2}, with the helicity weights that are of course consistent with the amplitude-anomaly definitions.
So any change in the conventions for the spinor-helicity properties will have to lead to the same overall phase either due to the polarization vectors in the expected result or due to the spinors in its on-shell counterpart, as those are the only sources for the little-group phase dependence and the related discrepancies in the on-shell ${\cal P}$ and ${\cal T}$ representations.

\section{Trivial Lorentz anomaly}
\label{app:Lorentz}

As discussed in \sec{sec:GaugeAnomaly}, anomalies are subject to the ambiguities inherent to the definition of the currents in QFT~\cite{Bardeen:1984pm,Bertlmann:1996xk}.
This is the reason for the different forms of the anomalies (\eg consistent and covariant anomalies).
However, anomalies are also subject to ambiguities due to the renormalization of the theory $\Gamma_{\text{ren}}$, since the local polynomial counterterms $\Gamma_\text{ct}$ may not be invariant under the anomalous symmetry
\begin{equation}
    {\cal A}(x)=\frac{\delta}{\delta\xi(x)}\delta_\xi\Gamma_{\text{ren}}=\frac{\delta}{\delta\xi(x)}\left(\delta_\xi\Gamma+\delta_\xi\Gamma_{\text{ct}}\right).
\end{equation}
Anomalies which are generated by a local polynomial, or in other words which can be canceled by $\Gamma_{\text{ct}}$, are artifacts of renormalization and are called trivial anomalies.
On the other hand, nontrivial anomalies are generated by the nonlocal part of $\Gamma$, and are regularization- and renormalization-independent.
In the case of gauge-symmetry anomalies, they truly violate the underlying theories.

Our bootstrap exhibits one such trivial anomaly in the pure-gauge sector~\eqref{eq:gALorentz2} of the Lorentz anomaly.
The first hint that this anomaly is trivial is that the coefficient $C_4$ has mass dimension $-2$, whereas all of the other anomalies that we bootstrapped have prefactors (after extracting the dimensionful coupling $\kappa$ associated with each graviton leg).
This implies that the field-theory operator stemming from \eqn{eq:gALorentz2} has mass dimension 6.

Let us consider the following local polynomial counterterm
\begin{equation}\!
\Gamma_{\text{ct}}=\!\int\!\dd^4x\sqrt{|g|}\,\omega_\la^{[\mu\nu]} R^\la{}_{\mu\rho\sigma}
   \big(a D^\rho F_\nu{}^\sigma\!+ b D^\rho \widetilde F_\nu{}^\sigma\big) ,
\end{equation}
where $\omega_\la^{[\mu\nu]}:=e_{\hat\imath}^\mu e_{\hat\jmath}^\nu \omega_\la^{\hat\imath\hat\jmath}$ and the coefficients $a$ and~$b$ have dimension $-2$.
Using $\delta^\text{L}_\xi \omega_{\la[\mu\nu]} = e^{\hat\imath}_\mu e^{\hat\jmath}_\nu \partial_\la\xi_{\hat\imath\hat\jmath}$, we find
\begin{widetext}
\begin{equation}
\delta^\text{L}_\xi\Gamma_{\text{ct}}=\!\int\!\dd^4x\sqrt{|g|}\xi_{\hat\imath\hat\jmath}\,\Bigg\{\!\Bigg(\frac{\partial^\la\!\sqrt{|g|}}{\sqrt{|g|}}R^{\hat\imath}{}_{\la\rho\sigma}+\partial^\la R^{\hat\imath}{}_{\la\rho\sigma}\!\Bigg)\big(a D^\rho F^{\hat\jmath\sigma}\!+ b D^\rho \widetilde F^{\hat\jmath\sigma}\big)+R^{\hat\imath}{}_{\la\rho\sigma}\big(a \partial^\la D^\rho F^{\hat\jmath\sigma}\!+ b \partial^\la D^\rho \widetilde F^{\hat\jmath\sigma}\big)\!\Bigg\}.
\label{eqapp:deltaGammact}
\end{equation}
\end{widetext}
Its contribution to the lowest-order on-shell Lorentz anomaly form-factor ${\cal A}^{[\hat\imath\hat\jmath]}_{gA}$ follows by taking one metric and one gauge-field variation of the integrand of \eqn{eqapp:deltaGammact} at $g=\eta$ and $A=0$, going to on-shell momentum space, and contracting with the polarization vectors.
Due to the transversality of the Christoffel symbols and Riemann tensor, only the second term of \eqn{eqapp:deltaGammact} contributes.
For example, for the term $a R^{\hat\imath}{}_{\la\rho\sigma} \partial^\la D^\rho F^{\hat\jmath\sigma}$ we find
\begin{align}\!\!
(p_{1\la} \varepsilon^{[\hat\imath}_1 - p^{[\hat\imath}_1 \varepsilon_{1\la})
(p_{1\rho}\varepsilon_{1\sigma} - p_{1\sigma}\varepsilon_{1\rho})\,
p^\la_2 p_2^\rho & (p_2^{\hat\jmath]} \varepsilon_2^\sigma - p_2^{\sigma}\varepsilon_2^{\hat\jmath]}) \nn \\*
\underset{3\,\text{pts}}{=} p_1^{[\hat\imath} p_2^{\hat\jmath]} (p_2\cdot\varepsilon_1)^2 & (p_1\cdot\varepsilon_2) .
\end{align}
Translating to spinors and combining this with the term $b R^{\hat\imath}{}_{\la\rho\sigma} \partial^\la D^\rho \tilde F^{\hat\jmath\sigma}$ involving the dual field strength, we finally obtain
\begin{equation}
\delta^\text{L}_\xi\Gamma_{\text{ct}}
 \underset{3\,\text{pts}}{\Rightarrow} -\tfrac{\kappa}{8\sqrt{2}} \times \Bigg\{\!
\begin{smallmatrix}
(a-ib) \braket{12}^4 |2]^{\dot\gamma} |2]^{\dot\delta} & \text{ for } (-,-) , \\
(a+ib) [12]^4 \bra{2}^\gamma \bra{2}^\delta & \text{ for } (+,+) ,\\
0 & \text{ for } (\pm,\mp) .
\end{smallmatrix}
\end{equation}
This has the same form as \eqn{eq:gALorentz2}, which proves that it is a trivial anomaly.

\bibliographystyle{apsrev4-1}
\bibliography{references}
\end{document}